\documentclass{aa}  

\bibpunct{(}{)}{;}{a}{}{,} 

\usepackage{graphicx}
\usepackage{txfonts}

\def\Msun{M_{\sun}}
\def\Lsun{L_{\sun}}

\def\Lstar{L_{\star}}
\def\Rstar{R_{\star}}
\def\Tstar{T_{\star}}
\def\ppt{\frac{\partial}{\partial t}}
\def\div#1{\nabla \! \cdot \! #1}
\def\grad#1{\nabla \! #1}
\def\kJ{\kappa^{g}_{\rm J}}
\def\kH{\kappa^{g}_{\rm H}}
\def\kS{\kappa^{g}_{\rm S}}
\def\Tg{T_{\rm g}}
\def\Td{T_{\rm d}}
\def\Tr{T_{\rm r}}
\def\Pg{P_{\rm g}}

\begin{document} 

   \title{Dynamic atmospheres and winds of cool luminous giants}

   \subtitle{I. Al$_2$O$_3$ and silicate dust in the close vicinity of M-type AGB stars}

   \author{S. Höfner \inst{1}
          \and
        S. Bladh \inst{2}
          \and
        B. Aringer \inst{2,3} 
          \and 
        R. Ahuja \inst{4}
          }

   \institute{Department of Physics and Astronomy, Division of Astronomy and Space Physics, 
                  Uppsala University, Box 516, SE-75120  Uppsala, Sweden 
              (\email{susanne.hoefner@physics.uu.se})
         \and
             Dipartimento di Fisica e Astronomia Galileo Galilei, Universit\`a di Padova, 
             Vicolo dell'Osservatorio 3, I-35122 Padova, Italy 
         \and
             Osservatorio Astronomico di Padova -- INAF, 
             Vicolo dell'Osservatorio 5, I-35122 Padova, Italy
         \and
             Department of Physics and Astronomy, Condensed Matter Theory Group, 
             Uppsala University, Box 516, SE-75120  Uppsala, Sweden
             }

   \date{Received ...; accepted ...}

   \abstract
   {In recent years, high spatial resolution techniques have given valuable insights into the complex atmospheres of AGB stars and their wind-forming regions. They allow to trace the dynamics of molecular layers and shock waves, to make estimates of dust condensation distances, and to obtain information on the chemical composition and size of dust grains close to the star. These are essential constraints for understanding the mass loss mechanism, which presumably involves a combination of atmospheric levitation by pulsation-induced shock waves and radiation pressure on dust, forming in the cool upper layers of the atmospheres. 
}
   {Spectro-interferometric observations indicate that Al$_2$O$_3$ condenses at distances of about 2 stellar radii or less, prior to the formation of silicates. Al$_2$O$_3$ grains are therefore prime candidates for producing the scattered light observed in the close vicinity of several M-type AGB stars, and they may be seed particles for the condensation of silicates at lower temperatures. The purpose of this paper is to study the necessary conditions for the formation of Al$_2$O$_3$ and the potential effects on mass loss, using detailed atmosphere and wind models. 
}
   {We have constructed a new generation of Dynamic Atmosphere \& Radiation-driven Wind models based on Implicit Numerics (DARWIN), including a time-dependent treatment of grain growth \& evaporation for both Al$_2$O$_3$ and Fe-free silicates (Mg$_2$SiO$_4$). The equations describing these dust species are solved in the framework of a frequency-dependent radiation-hydrodynamical model for the atmosphere \& wind structure, taking pulsation-induced shock waves and periodic luminosity variations into account.
}
   {Condensation of Al$_2$O$_3$ at the close distances and in the high concentrations implied by observations requires high transparency of the grains in the visual and near-IR region to avoid destruction by radiative heating. We derive an upper limit for the imaginary part of the refractive index $k$ around $10^{-3}$ at these wavelengths. For solar abundances, radiation pressure due to Al$_2$O$_3$ is too low to drive a wind. Nevertheless, this dust species may have indirect effects on mass loss. The formation of composite grains with an Al$_2$O$_3$ core and a silicate mantle can give grain growth a head start, increasing both mass loss rates and wind velocities. Furthermore, our experimental core-mantle grain models lead to variations of visual and near-IR colors during a pulsation cycle which are in excellent agreement with observations. 
}
   {Al$_2$O$_3$ grains are promising candidates for explaining the presence of gravitationally bound dust shells close to M-type AGB stars, as implied by both scattered light observations and mid-IR spectro-interferometry. The required level of transparency at near-IR wavelengths is compatible with impurities due to a few percent of transition metals (e.g. Cr), consistent with cosmic abundances. Grains consisting of an Al$_2$O$_3$ core and an Fe-free silicate mantle, with total grain radii of about 0.1-1 micron, may be more efficient at driving winds by scattering of stellar photons than pure Fe-free silicate grains. 
}


   \titlerunning{}

   \authorrunning{}

   \maketitle


\section{Introduction}

The recent progress in high spatial resolution techniques, spanning wavelengths from the visual to the radio regime, is leading to a wealth of new information on dynamical atmospheres of AGB stars and on dust formation in the close vicinity of these objects. Striking examples are detections of significant deviations from spherical symmetry in the photospheric and dust-forming layers \citep[see, e.g.,][]{haubois15,stew16,ohnaka16} which are probably due to large-scale convective motions as predicted by 3D 'star-in-a-box' models \citep[][]{frey08,witt16}. In addition to general insights on photospheric and circumstellar structures, high spatial resolution techniques can provide specific measurements which are essential for our understanding of wind mechanisms, in particular condensation distances of various dust species \citep[e.g.][]{ireland05,witt07,zhao11,zhao12,karo13,sacu13} and grain sizes in the wind acceleration zone \citep[e.g.][]{norris12,ohnaka16}. 
 
The basic scenario for the mass loss of AGB stars involves a Pulsation-Enhanced Dust-DRiven Outflow \citep[abbreviated as PEDDRO below; see][for a recent review on this topic]{hoefner15}.
Stellar pulsations (probably in combination with large-scale convective motions) trigger strong shock waves which propagate outwards through the atmosphere. These shocks compress the gas and, intermittently, lift it to distances from the stellar photosphere where temperatures are sufficiently low to allow for dust condensation. The solid particles gain momentum by absorbing and scattering stellar photons, and are accelerated away from the star. Through frequent collisions with atoms and molecules the dust grains transfer momentum to the gas, triggering a wind. 

The PEDDRO scenario is supported by various types of observations, not the least by high-resolution spectroscopy which allows to trace gas velocities from the photosphere to the wind region via Doppler shifts in line profiles \citep[e.g.][]{HHR82,SW00,nowo10}. Nevertheless, some open questions remain. In particular, there is an on-going debate on the nature of the wind-driving grains in M-type AGB stars. Magnesium-iron silicates (olivine- and pyroxene-type materials) seem to be obvious candidates, given the relatively high abundances of their constituent elements (Si, Mg, Fe and O) and the prominence of silicate features in mid-IR spectra of circumstellar dust shells. Detailed models, however, show that silicates have to be basically Fe-free in layers close to the star where the wind originates, in order to avoid destruction by radiative heating \citep[][]{woit06b}. The resulting low levels of absorption at visual and near-IR wavelengths lead to a low radiative pressure, which is insufficient to drive an outflow. 

In a series of papers based on detailed atmosphere \& wind models, we have therefore investigated an alternative mechanism, i.e. wind-driving by scattering of stellar photons on Fe-free silicate grains \citep[][]{hoefner08,bladh13,bladh15}. This scenario requires grain sizes of about 0.1 -- 1 micron, in order to make the dust particles efficient at scattering radiation in the near-IR wavelength region where the stellar flux peaks. Earlier, there was doubt whether grains could grow that big in stellar atmospheres. In recent years, however, there is increasing observational evidence for the existence of such large particles in the close vicinity of AGB stars and supergiants \citep[e.g.][]{norris12,scicluna15,ohnaka16}. Furthermore, detailed atmosphere and wind models based on this mechanism show good agreement with observations regarding mass loss rates and wind velocities, as well as visual and near-IR colors, and their variation with pulsation phase \citep[][]{bladh13,bladh15}. 

In the present paper we take these studies a step further, including Al$_2$O$_3$ as an additional dust species in our models. An isotopic analysis of presolar Al$_2$O$_3$ grains by \citet[][]{nittler97} suggests an origin of these grains in winds of AGB stars. Spectro-interferometric observations indicate that this material forms at distances of about 2 stellar radii or less, prior to silicate condensation \citep[e.g.][]{witt07,zhao11,zhao12,karo13}. Al$_2$O$_3$ has been discussed as a possible alternative to silicates as a source of the scattered light observed close to several AGB stars \citep[e.g.][]{ireland05,norris12,ohnaka16}, and as potential seed particles for the condensation of silicates further out in the atmosphere where lower temperatures prevail \citep[e.g.][]{KoSo97a,KoSo97b}. 

With our new generation of Dynamic Atmosphere and Radiation-driven Wind models based on Implicit Numerics (DARWIN) we investigate these questions, as well as other issues related to the formation of Al$_2$O$_3$ and its possible effects on mass loss. In addition to a time-dependent description of dust formation \& destruction, these radiation-hydrodynamical models feature frequency-dependent radiative transfer (including gas and dust opacities), pulsation-induced atmospheric shock waves and periodic luminosity variations. A detailed description of the DARWIN code is given in Sect.~\ref{s_methods}, and the modelling results are presented in Sect.~\ref{s_results}, followed by a discussion and comparison with observations in Sect.~\ref{s_discussion}, and a short summary of results \& conclusions in Sect.~\ref{s_conclusions}.

\section{The DARWIN code}\label{s_methods}

The DARWIN models presented in this paper describe the time-dependent structure of the stellar atmosphere and wind (i.e. velocities, densities, temperatures, dust properties, etc., as a function of time and radial distance from the stellar center), as determined by the coupled system of gas dynamics, radiative processes and non-equilibrium dust formation. The pulsating AGB star is characterized by its fundamental parameters, i.e., mass, luminosity, effective temperature and chemical composition, as well as period and amplitude of the pulsation. Dependent on these parameters, the models give two types of results, (i) mass loss rates, wind velocities \& dust yields, as direct output of the DARWIN code, and (ii) spectra, light curves, visibilities, and other synthetic observables, computed {\it a posteriori} from snapshots of the radial structures using the COMA code \citep[][]{BADiss00,aringer09}.

The spherically symmetric models cover a region with an inner boundary below the stellar photosphere but above the driving zone of the pulsations. The physical description is optimized for the optically thin atmospheres and winds with their complex molecular and dust chemistry (dominated by radiative energy transport and strong radiating shocks), in contrast to the optically thick, convective stellar interior where the pulsations are excited. The effects of stellar pulsation are simulated by temporal variations of gas velocity and luminosity at the inner boundary of the model. The location of the outer boundary depends on the emerging dynamics: for models which form stellar winds it is typically at a distance of 20--30 stellar radii (allowing outflow), for models without a wind the outer boundary is close to the photosphere, following the quasi-ballistic movements of the upper atmospheric layers. The computations start with a hydrostatic, dust-free, atmospheric structure corresponding to the fundamental parameters of the star. The effects of pulsation are introduced gradually by increasing the amplitude up to the full value over typically tens of periods and the simulations are run for several hundred pulsation periods to avoid transient effects.

The DARWIN models presented here build on earlier generations of our non-grey, time-dependent atmosphere \& wind models \citep[see][]{hoefner_etal03, hoefner08,bladh15}. For the model components which have been described in detail before we only give a short overview in this paper (with equations and more details presented in appendices). The modeling of non-equilibrium dust condensation for stars with C/O $<$ 1, however, is described in some detail in this section, in particular regarding wind-driving silicate grains (Mg$_2$SiO$_4$) and the newly-implemented species Al$_2$O$_3$.

\subsection{Dust: grain growth and composition}\label{s_meth_dust}

In atmospheres and winds of AGB stars grain growth usually proceeds far from chemical equilibrium. One reason is the critical influence of the stellar radiation, resulting in grain temperatures that differ from the gas temperature, and that are strongly dependent on the composition and optical properties of the dust particles. Furthermore, at distances from the stellar photosphere where temperatures are low enough to allow for dust condensation, low gas densities lead to slow grain growth, on timescales which are comparable to those of stellar pulsation and ballistic atmospheric motions. Consequently, the grains do not adjust instantaneously to the prevailing local conditions, as in thermal and chemical equilibrium, but their properties are dependent on the preceding evolution of the layers in which they are embedded. Also, if dust condensation triggers an outflow, rapidly decreasing densities in the wind will efficiently quench additional grain growth, often resulting in condensation degrees distinctly below unity for chemical elements building up the wind-driving dust grains.

To take these processes into account, the DARWIN models use a time-dependent kinetic description of grain growth which can be summarized in the following way: The grains grow by addition of abundant atoms and molecules from the gas phase to the surface of the solid particles. In contrast to other environments (e.g. proto-planetary discs), grain-grain collisions and coagulation can be neglected due to the combination of low densities and short dynamical timescales. The dust grains are characterized by their particle radius, and their size-dependent optical properties are calculated using Mie theory and refractive index data corresponding to their chemical composition. 
Since dust nucleation (i.e. the initial formation of condensation nuclei from the gas phase) is still poorly understood for stars with C/O$\,<\,$1 \citep[see, e.g.,][and reference therein]{gobrecht16,gail16arx}, we assume that tiny seed particles (with sizes corresponding to 1000 monomers, i.e. basic building blocks of the solid) exist prior to the onset of grain growth, and we parameterize them by their abundance relative to hydrogen. A more detailed description of the dust model used here for the close circumstellar environment of M-type AGB stars is given in the following subsections. Corresponding information for models of stars with C/O $>$ 1 can be found in earlier papers \citep[see, e.g,][and references therein]{hoefner_etal03,matt10,matt11,erik14}.

\subsubsection{Silicates}\label{s_meth_ol}

At distances of a few stellar radii, where observations place the dust condensation zone around AGB stars, dust grains are subject to substantial radiative heating. The inclusion of Fe in silicate grains (which a purely kinetic picture would favor due to similar abundances of Mg and Fe in a solar mixture), even in small amounts, will heat the particles above the sublimation temperature  \citep[e.g.,][]{woit06b,BH12,bladh15}. Therefore we assume here that the silicates in this region are basically Fe-free. We consider the formation of Fe-free olivine-type silicates, i.e. Mg$_2$SiO$_4$, 
from abundant molecules in the gas phase according to the net reaction
\begin{equation}\label{e_path_ol}
  {\rm 2 \, Mg + SiO + 3 \, H_2 O }  \,\,  \longrightarrow  \,\,  {\rm Mg_2 SiO_4 + 3 \, H_2 } \, .
\end{equation}
Basically, the description of dust condensation and evaporation used here follows the method presented by \citet[]{GS99}, however, with one decisive difference due to the Fe-free nature of the grains: In a solar element mixture the abundance of Si and Mg are comparable and the abundance of SiO will be determined by the abundance of Si in the gas phase. Since 2~Mg atoms are required for each SiO added to the dust particles, Mg will be the limiting factor for grain growth under these circumstances (with H$_2$O being much more abundant than either Mg or SiO). Consequently, the equation describing the time evolution of the grain radius $a_{\rm sil}$, accounting for grain growth and decomposition, can be formulated as
\begin{eqnarray}\label{e_rate_ol}
  \frac{d a_{\rm sil}}{d t} 
    & = & V_{\rm sil} \, \left[  J^{\rm gr}_{\rm sil}  - J^{\rm dec}_{\rm sil} \right] \nonumber \\
    & = & \frac{1}{2} \, V_{\rm sil} \, \alpha_{\rm Mg} v_{\rm Mg} n_{\rm Mg}
            \left[ \, 1 - \frac{p_{\rm v , Mg}}{n_{\rm Mg} k T_g} \sqrt{\frac{T_g}{T_d}} \, \right]
\end{eqnarray}
where $V_{\rm sil} = A_{\rm sil} \, m_{\rm H} / \rho_{\rm sil}$ is the volume of the nominal monomer (atomic weight $A_{\rm sil}=140$ for Mg$_2$SiO$_4$, $m_{\rm H}$ = hydrogen mass) 
and $\rho_{\rm sil}$ is the bulk density of the material. 
The growth and decomposition rates are given by 
\begin{eqnarray}
  J^{\rm gr}_{\rm sil}   & = & \frac{1}{2} \, \alpha_{\rm Mg} v_{\rm Mg} n_{\rm Mg} 
                                         \label{e_jgr_ol} \\
  J^{\rm dec}_{\rm sil} & = & \frac{1}{2} \, \alpha_{\rm Mg} v_{\rm Mg} \frac{p_{\rm v , Mg}}{k T_g} \sqrt{\frac{T_g}{T_d}}
\end{eqnarray}
respectively (accounting for different gas and dust temperatures, $T_g$ and $T_d$). The root mean square thermal velocity of the Mg atoms in the gas phase is given by $v_{\rm Mg} = \sqrt{k T_g / 2 \pi m_{\rm Mg}}$, the symbol $n_{\rm Mg}$ denotes the number density of Mg atoms in the gas phase (taking depletion of Mg due condensation into account) and $\alpha_{\rm Mg}$ is the sticking coefficient (assumed to be 1, considering that the Fe-free silicate grains are much cooler than the surrounding gas). The quantity $p_{\rm v , Mg}$ appearing in the decomposition rate represents the partial pressure of Mg in chemical equilibrium between the gas phase and Mg$_2$SiO$_4$, according to the net reaction in Eq.\,(\ref{e_path_ol}) and its reverse process. In our models it is calculated based on the data for free energies by \citet[]{SH90}.
The bulk density is $\rho_{\rm sil} = 3.27$ [g/cm$^3$], consistent with the optical data for amorphous Mg$_2$SiO$_4$ by \citet[][see Sect.~\ref{s_nk_cor}]{jaeger03}. This choice of data seems natural since observations of AGB stars indicate that amorphous silicates are more abundant than their crystalline counterparts \citep[see, e.g.,][]{devries10}.

\begin{figure}
\centering
\includegraphics[width=\hsize]{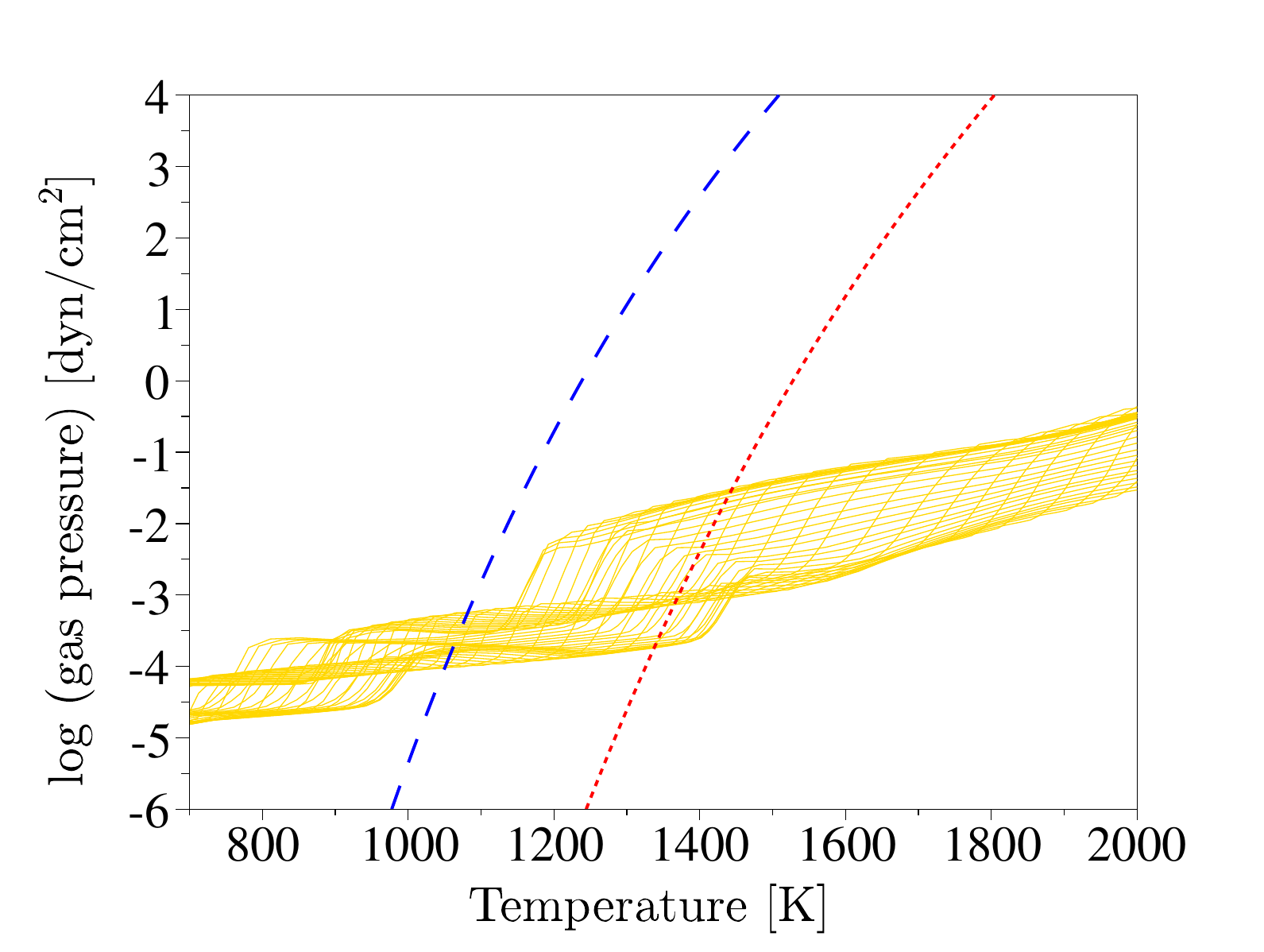}
     \caption{Stability limits for Mg$_2$SiO$_4$ (dashed) and Al$_2$O$_3$ (dotted), see text for details. Snapshots of pressure--temperature structures for a typical dynamical model (An315u3, see Tab.~\ref{t_mod}) are shown in yellow.
                   }
      \label{f_pstl}
\end{figure}

Depending on the relative values of the growth and decomposition rates, the r.h.s. of Eq.~(\ref{e_rate_ol}) can be positive or negative, describing growing or shrinking grain radii. In thermal and chemical equilibrium between the gas and dust phases the two rates balance each other, and the grain size is constant. In this case, the stability limit of a material can be described as the temperature (for a given gas pressure) above which all material is in the gas phase, and none condensed into a solid state. Fig.~\ref{f_pstl} shows this stability limit for Mg$_2$SiO$_4$ and Al$_2$O$_3$, compared to snapshots of the variable pressure-temperature structure of a typical model, indicating condensation temperatures for the two dust species of about 1100\,K and 1400\,K, respectively.

\begin{figure}
\centering
\includegraphics[width=\hsize]{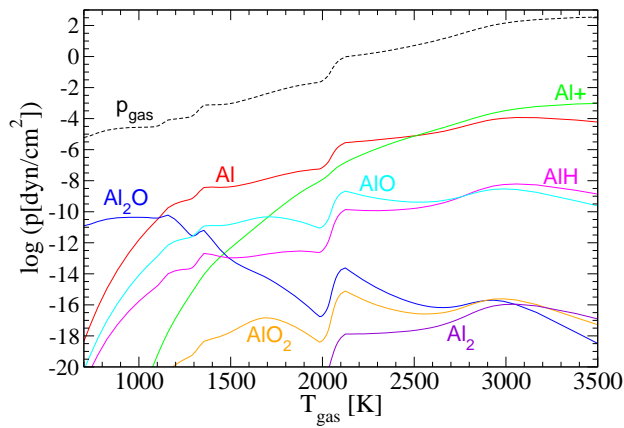}
      \caption{Partial pressures of Al-bearing atomic and molecular species versus gas temperature for a snapshot taken from a typical model (An315u3, see Tab.~\ref{t_mod}). The values are computed assuming chemical equilibrium in the gas phase.
                   }
      \label{f_Al_gas}
\end{figure}

\subsubsection{Al$_2$O$_3$}\label{s_meth_crn}

Regarding the formation of Al$_2$O$_3$ we follow the reasoning by \citet[]{GS14}, assuming that Al predominantly exists as free atoms in the gas phase in the relevant temperature range (about 1400--1100~K, i.e. below the condensation temperature of Al$_2$O$_3$ but above that of Mg$_2$SiO$_4$, see Fig.~\ref{f_pstl}). This assumption is consistent with the partial pressures derived for various Al-bearing species in our models (see Fig.~\ref{f_Al_gas}), leading to the net reaction 
\begin{equation}\label{e_path_crn}
  {\rm 2 \, Al + 3 \, H_2 O }  \,\,  \longrightarrow  \,\,  {\rm Al_2 O_3 + 3 \, H_2 } \, .
\end{equation}
Since H$_2$O is much more abundant than Al in a solar mixture, the growth rate will be limited by the addition of 2 Al atoms per monomer. Accordingly, the equation describing the growth and decomposition of Al$_2$O$_3$ is
\begin{eqnarray}\label{e_rate_crn}
  \frac{d a_{\rm alu}}{d t} 
      & = & V_{\rm alu} \, \left[  J^{\rm gr}_{\rm alu}  - J^{\rm dec}_{\rm alu} \right]  \nonumber \\
      & = & \frac{1}{2} \, V_{\rm alu} \, \alpha_{\rm Al} v_{\rm Al} n_{\rm Al}
            \left[ \, 1 - \frac{p_{\rm v ,Al}}{n_{\rm Al} k T_g} \sqrt{\frac{T_g}{T_d}} \, \right]
\end{eqnarray}
where $a_{\rm alu}$ is the radius of the grains, $V_{\rm alu} = A_{\rm alu} \, m_{\rm H} / \rho_{\rm alu}$ is the volume of the nominal monomer (atomic weight $A_{\rm alu}=102$, bulk density $\rho_{\rm alu} = 3.99$ [g/cm$^3$] for Al$_2$O$_3$) and the growth and decomposition rates are given by 
\begin{eqnarray}
  J^{\rm gr}_{\rm alu}   & = & \frac{1}{2} \, \alpha_{\rm Al} v_{\rm Al} n_{\rm Al} 
                                          \label{e_jgr_crn} \\
  J^{\rm dec}_{\rm alu} & = & \frac{1}{2} \, \alpha_{\rm Al} v_{\rm Al} \frac{p_{\rm v , Al}}{k T_g} \sqrt{\frac{T_g}{T_d}}
\end{eqnarray}
respectively. The root mean square thermal velocity of the Al atoms in the gas phase is $v_{\rm Al} = \sqrt{k T_g / 2 \pi m_{\rm Al}}$. The quantity $n_{\rm Al}$ denotes the number density of Al atoms in the gas phase (accounting for depletion due to condensation) and $\alpha_{\rm Al}$ the sticking coefficient (assumed to be 1, based on grain temperatures being much lower than those of the surrounding gas). The symbol $p_{\rm v , Al}$ appearing in the decomposition rate represents the partial pressure of Al in chemical equilibrium between the gas phase and Al$_2$O$_2$, according to the net reaction in Eq.\,(\ref{e_path_crn}) and its reverse process \citep[calculated using free energies by][]{SH90}.  

In terms of the kinetic grain growth model used here, crystalline and amorphous phases of a dust species are distinguished mainly by their bulk density and optical properties, as indicated above. 
An important point of this paper is to test a range of options for the absorption properties of Al$_2$O$_3$ in the visual and near-IR regime, which depend strongly on impurities, i.e., inclusions of transition metals (see Sect.~\ref{s_nk_cor} for details). In view of the high intrinsic uncertainties 
of the visual \& near-IR optical data (which determines the impact of radiative heating on grain temperatures) it is difficult to make reliable predictions if crystalline or amorphous material is more likely to form in AGB star atmospheres. Mid-IR spectral features of circumstellar dust as well as presolar dust grains found in meteorites indicate, however, that both amorphous and cystalline grains of Al$_2$O$_3$ are produced in AGB stars \citep[see, e.g.,][and references therein]{stroud04}. 
In this paper we use mid-IR optical data for amorphous Al$_2$O$_3$ \citet[][data set x2]{bege97} which is a common choice in the literature.

\subsection{Gas dynamics and radiative transfer}

As discussed above, dust formation depends critically on the ambient thermodynamical conditions (temperatures, densities and abundances of relevant atoms and molecules), which in turn are strongly affected by gas dynamics and radiative processes. Therefore, in order to obtain realistic results, dust formation has to be put into the context of a detailed dynamical model which takes these various feedback mechanisms into account.
 
The time-dependent structures of the atmospheres and winds are governed by the basic conservation laws of mass, momentum and energy for the strongly coupled system of gas, dust and radiation. More specifically, to obtain the models presented in this paper, we solve the equations of hydrodynamics (equation of continuity, equation of motion and energy equation) together with the frequency-integrated zeroth and first moment equations of the radiative transfer equation (accounting for the energy and momentum balance of the radiative field). The equation of motion takes external forces due to gravity and radiative pressure into account, in addition to gas pressure gradients. The equation describing the internal energy of the gas component includes radiative heating and cooling, as well as changes due to compression and expansion of the gas, caused by external forces. A detailed description of these equations is given in App.~\ref{app_rhd}. The variable boundary condition which are used to simulate effects of stellar pulsation are discussed in App.~\ref{app_inbc}, and the numerical methods are briefly summarized in App.~\ref{app_num}.

The system of hydrodynamical equations requires a closing condition in the form of an equation of state for the gas, connecting the thermal pressure and the temperature of the gas with its density and specific internal energy. To keep the models comparable to previous results we use the same assumptions about the equation of state as in earlier papers, i.e. a perfect gas with $\gamma = 5/3$ and $\mu = 1.26$. The good agreement of the resulting spectra with observations can be taken as an indication that this assumption, despite its simplicity, leads to reasonably realistic atmospheric structures. 

A correct modeling of radiative heating and cooling, which strongly affects the temperatures of both gas and dust, requires a frequency-dependent treatment of radiative transfer. The DARWIN code alternates between computing the dynamical structures (based on the conservation laws discussed above) and solving the frequency-dependent radiative transfer equation (with these structures as input) to obtain proper frequency means of gas and dust opacities which are needed for solving the radiation-hydrodynamical equations, in turn. 
The models presented in this paper use 319 frequency points for this procedure, distributed over a wavelength range between 0.25 and 25 $\mu$m. 
The gas opacities are pre-tabulated for given sets of elemental abundances using the COMA code 
\citep[][]{BADiss00,aringer09}. For the solar composition models discussed here we adopted the values from \citet[][]{AndGre89}, except for C, N and O where we took the data from \citet[][]{GrevSauv94}. 

In contrast to the gas, the dust opacities are calculated on the fly, using the current grain sizes $a_{\rm gr}$ in each layer, and refractive index data corresponding to the chemical composition of the grains (the optical data used in this paper is discussed in Sect.~\ref{s_results}). The opacity relevant for the radiation pressure can be written as
\begin{equation}
  \kappa_{\rm acc} (a_{\rm gr},\lambda) = \frac{3 \, A_{\rm mon}}{4 \, \rho_{\rm gr}} \, 
    \frac{Q_{\rm acc} (a_{\rm gr},\lambda)}{a_{\rm gr}} \,
    \frac{\varepsilon_c \, f_c}{s \, (1 + 4 \,  \varepsilon_{\rm He})} 
\end{equation}
where $A_{\rm mon}$ and $\rho_{\rm gr}$ are the atomic weight of the monomer and the bulk density of the grain material, respectively (see Sect.~\ref{s_meth_dust}), $\varepsilon_{\rm He}$ is the abundance of He, and $\varepsilon_c$, $f_c$ and $s$ denote the abundance, condensation fraction and stochiometric coefficient of a critical element, defining the amount of dust \citep[see, e.g.,][for a discussion]{BH12}. The efficiency factor $Q_{\rm acc} (a_{\rm gr},\lambda)$, defined as the ratio of radiative cross section to geometrical cross section of a grain, contains contributions from true absorption and scattering, i.e.,
\begin{equation}
  Q_{\rm acc}  =  Q_{\rm abs} + ( 1 -  g_{\rm sca} ) \, Q_{\rm sca} 
\end{equation}
where $g_{\rm sca}$ is an asymmetry factor describing deviations from isotropic scattering \citep[see, e.g.,][]{kruegel03}. The efficiency factors and $g_{\rm sca}$ are computed using Mie theory (program {\tt BHMIE} by B.T.~Draine\footnote{{\tt www.astro.princeton.edu/\~{ }draine/scattering.html}}). 
The opacity defining radiative heating and cooling (and therefore the grain temperature, see App.~\ref{app_rhd}) can be written in a similar general form as $\kappa_{\rm acc}$, replacing the efficiency factor $Q_{\rm acc}$ with the corresponding quantity for true absorption, i.e. $Q_{\rm abs}$, since pure scattering of photons does not change the energy of a dust grain.

\begin{table*}
\caption{\label{t_mod}DARWIN models with outflows driven by scattering of stellar photons on Fe-free silicate grains. The letters A and B denote 2 different combinations of stellar parameters (see text), $n_d/n_{\rm H}$  the seed particle abundance, and $\Delta u_{\rm P}$ the velocity amplitude at the inner boundary. The resulting wind and dust properties listed here are temporal means of the mass loss rate $\dot{M}$, the wind velocity $u_{\rm ext}$, the fraction of Si condensed into grains $f_{\rm Si}$, the fraction of Mg condensed into grains $f_{\rm Mg}$, and the grain radius $a_{gr}$ at the outer boundary. In this set of models Al$_2$O$_3$ is included as a separate, passive dust species (ignoring its feedback on the atmosphere \& wind). Condensation fractions of Al, denoted by $f_{\rm Al}$, are given for 3 different cases of radiative heating, discussed in Sect.~\ref{s_crn_cond}. $T_{\rm alu}$ and $T_{\rm sil}$ are the grain temperatures of Al$_2$O$_3$ and Mg$_2$SiO$_4$, respectively. 
}
\centering
\begin{tabular}{l|cc|ccccc|ccc}
\hline\hline
  & & & & & & & & & \\
  model &  $n_d/n_{\rm H}$  &  $\Delta u_{\rm P}$ &  $\dot{M}$  & $u_{\rm ext}$  & $f_{\rm Si}$  & $f_{\rm Mg}$ & $a_{gr}$  &  $f_{\rm Al}$  & $f_{\rm Al}$  & $f_{\rm Al}$ \\
  name  &  & [km/s] & [$\Msun$/yr] & [km/s] &  &  & [$\mu$m] & $T_{\rm alu}$\,=\,$T_{\rm sil}$ & {\em high k} & {\em low k}\\ 
  & & & & & & & & & \\
\hline
  & & & & & & & & & \\
  An115u3 & $1.0 \cdot 10^{-15}$ & 3.0 & $1 \cdot 10^{-7}$ & 2 & 0.17 & 0.32 & 0.47 & 0.29 & 0.04 & 0.95 \\
  An315u3 & $3.0 \cdot 10^{-15}$ & 3.0 & $3 \cdot 10^{-7}$ & 5 & 0.20 & 0.37 & 0.34 & 0.41 & 0.07 & 1.00 \\
  An114u3 & $1.0 \cdot 10^{-14}$ & 3.0 & $4 \cdot 10^{-7}$ & 7 & 0.27 & 0.50 & 0.25 & 0.60 & 0.14 & 1.00 \\
  An314u3 & $3.0 \cdot 10^{-14}$ & 3.0 & $4 \cdot 10^{-7}$ & 7 & 0.38 & 0.71 & 0.20 & 0.86 & 0.36 & 1.00 \\
  & & & & & & & & & \\
  An115u4 & $1.0 \cdot 10^{-15}$ & 4.0 & $3 \cdot 10^{-7}$ & 3 & 0.17 & 0.32 & 0.47 & 0.55 & 0.19 & 0.98\\
  An315u4 & $3.0 \cdot 10^{-15}$ & 4.0 & $5 \cdot 10^{-7}$ & 6 & 0.20 & 0.37 & 0.34 & 0.78 & 0.44 & 1.00\\
  An114u4 & $1.0 \cdot 10^{-14}$ & 4.0 & $6 \cdot 10^{-7}$ & 9 & 0.27 & 0.50 & 0.25 & 0.84 & 0.59 & 1.00 \\
  An314u4 & $3.0 \cdot 10^{-14}$ & 4.0 & $5 \cdot 10^{-7}$ & 8 & 0.37 & 0.69 & 0.20 & 0.96 & 0.82 & 1.00 \\
  & & & & & & & & & \\
  Bn316u3 & $3.0 \cdot 10^{-16}$ & 3.0 & $3 \cdot 10^{-7}$ & 2 & 0.13 & 0.24 & 0.64 & 0.98 & 0.77 & 1.00 \\
  Bn115u3 & $1.0 \cdot 10^{-15}$ & 3.0 & $5 \cdot 10^{-7}$ & 5 & 0.14 & 0.26 & 0.44 & 0.99 & 0.94 & 1.00 \\
  Bn315u3 & $3.0 \cdot 10^{-15}$ & 3.0 & $7 \cdot 10^{-7}$ & 7 & 0.17 & 0.32 & 0.33 & 1.00 & 0.99 & 1.00 \\
  Bn114u3 & $1.0 \cdot 10^{-14}$ & 3.0 & $8 \cdot 10^{-7}$ & 8 & 0.24 & 0.45 & 0.24 & 1.00 & 1.00 & 1.00 \\
  Bn314u3 & $3.0 \cdot 10^{-14}$ & 3.0 & $8 \cdot 10^{-7}$ & 8 & 0.35 & 0.65 & 0.19 & 1.00 & 1.00 & 1.00 \\
  & & & & & & & & & \\
  Bn316u4 & $3.0 \cdot 10^{-16}$ & 4.0 & $3 \cdot 10^{-7}$ & 2 & 0.13 & 0.24 & 0.64 & 0.99 & 0.62 & 1.00 \\
  Bn115u4 & $1.0 \cdot 10^{-15}$ & 4.0 & $5 \cdot 10^{-7}$ & 4 & 0.13 & 0.24 & 0.43 & 1.00 & 0.86 & 1.00 \\
  Bn315u4 & $3.0 \cdot 10^{-15}$ & 4.0 & $7 \cdot 10^{-7}$ & 6 & 0.17 & 0.32 & 0.33 & 1.00 & 0.97 & 1.00 \\
  Bn114u4 & $1.0 \cdot 10^{-14}$ & 4.0 & $8 \cdot 10^{-7}$ & 8 & 0.24 & 0.45 & 0.24 & 1.00 & 0.99 & 1.00 \\
  Bn314u4 & $3.0 \cdot 10^{-14}$ & 4.0 & $8 \cdot 10^{-7}$ & 8 & 0.34 & 0.64 & 0.19 & 1.00 & 1.00 & 1.00 \\
  & & & & & & & & & \\
\hline
\end{tabular}
\tablefoot{The model names are constructed in the following way: The capital letter (A, B) represents a combination of stellar parameters (see Sect.~\ref{s_results}), the letter n followed by a 3-digit number stands for the seed particle abundance (i.e. n316 for $n_d/n_{\rm H} = 3 \cdot 10^{-16}$, n115 for $n_d/n_{\rm H} = 1 \cdot 10^{-15}$, etc.) and the letter u followed by 3 or 4 represents the pulsation amplitude (km/s).}
\end{table*}

\begin{figure}
\centering
\includegraphics[width=\hsize]{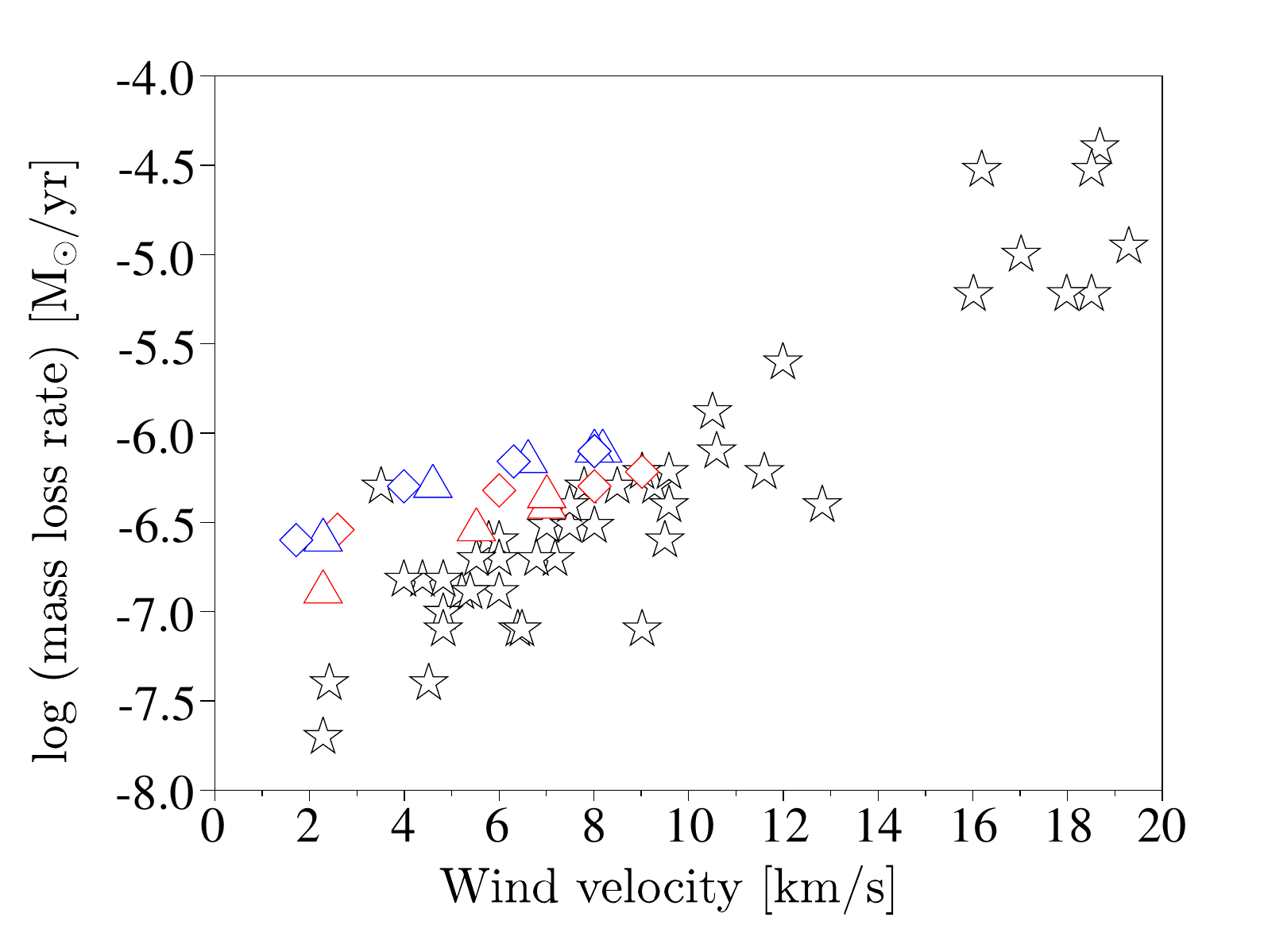}
     \caption{Mass loss rate versus wind velocity for M-type AGB stars: black symbols (stars) mark observations by \citet{olof02} and \citet{gonz03}, red symbols models of series A and blue symbols models of series B (triangles and squares indicate pulsation amplitudes of 3 and 4 km/s, 
respectively; see Tab.~\ref{t_mod} for details). Note that the stellar parameters of the models where chosen to produce weak to moderate winds, as such environments seem to be favorable for 
Al$_2$O$_3$ formation.
                   }
      \label{f_dMdt_v_oli}
\end{figure}

\section{Model parameters and results}\label{s_results}

Mid-IR spectra of circumstellar envelopes show distinct features of silicate dust for AGB stars with more massive winds, while Al$_2$O$_3$ features seem to be more pronounced in objects with low mass loss rates \citep[see, e.g.,][]{LoMaPom00,sloan03a,karo13}. Since the main purpose of this paper is to study the formation of Al$_2$O$_3$ and its interplay with silicate dust, we choose 2 sets of stellar parameters for the hydrostatic initial models (labeled A and B) which result in low to moderate mass loss rates and wind velocities:  

\begin{center}
\begin{tabular}{l|ccc}
\hline\hline
      & $M_{\ast}$  [$\Msun$] & $L_{\ast}$ [$\Lsun$] & $T_{\ast}$ [K] \\
\hline
   A & 1 & 5000 & 2800 \\
   B & 1 & 7000 & 2700 \\
\hline
\end{tabular}
\end{center}
Model A is assigned a pulsation period of 310 days and models B 390 days, based on the period-luminosity relation by \citet{feast89}. For each of the 2 models we use 2 pulsation amplitudes, i.e. velocity amplitudes at the inner boundary $\Delta u_{\rm P} = 3\,$[km/s] and $\Delta u_{\rm P} = 4\,$[km/s]. In combination with $f_L = 2$ (see App.~\ref{app_inbc}) this leads to bolometric amplitudes typical of AGB stars. For each of these 4 configurations of stellar and pulsation parameters, the seed particle abundance $n_d/n_{\rm H}$, i.e. the ratio of the number densities of dust grains and H nuclei, is varied over a wide range (up to 2 orders of magnitude), covering values which, in principle, allow grains to reach sizes necessary for driving a wind by photon scattering (about $0.1 - 1$ micron). 

The presentation of modelling results regarding Al$_2$O$_3$ consists of two major parts: First, we discuss the necessary conditions for the condensation of this species in the vicinity of AGB stars (using DARWIN models where Al$_2$O$_3$ grains are treated as a passive species in winds driven by silicate grains, Sect.~\ref{s_crn_cond}). Then we describe its effects on atmospheres and winds (based on DARWIN models featuring Al$_2$O$_3$ grains only, or composite grains consisting of Al$_2$O$_3$ cores with silicate mantles, Sect.~\ref{s_crn_eff}). Before discussing Al$_2$O$_3$ formation and its possible consequences for atmospheric dynamics, however, we need to re-examine the role of Mg$_2$SiO$_4$ as a wind driver.

\subsection{Effects of revised growth rates for silicate grains}\label{s_rates}

\begin{figure*}
\centering
\includegraphics[width=\hsize]{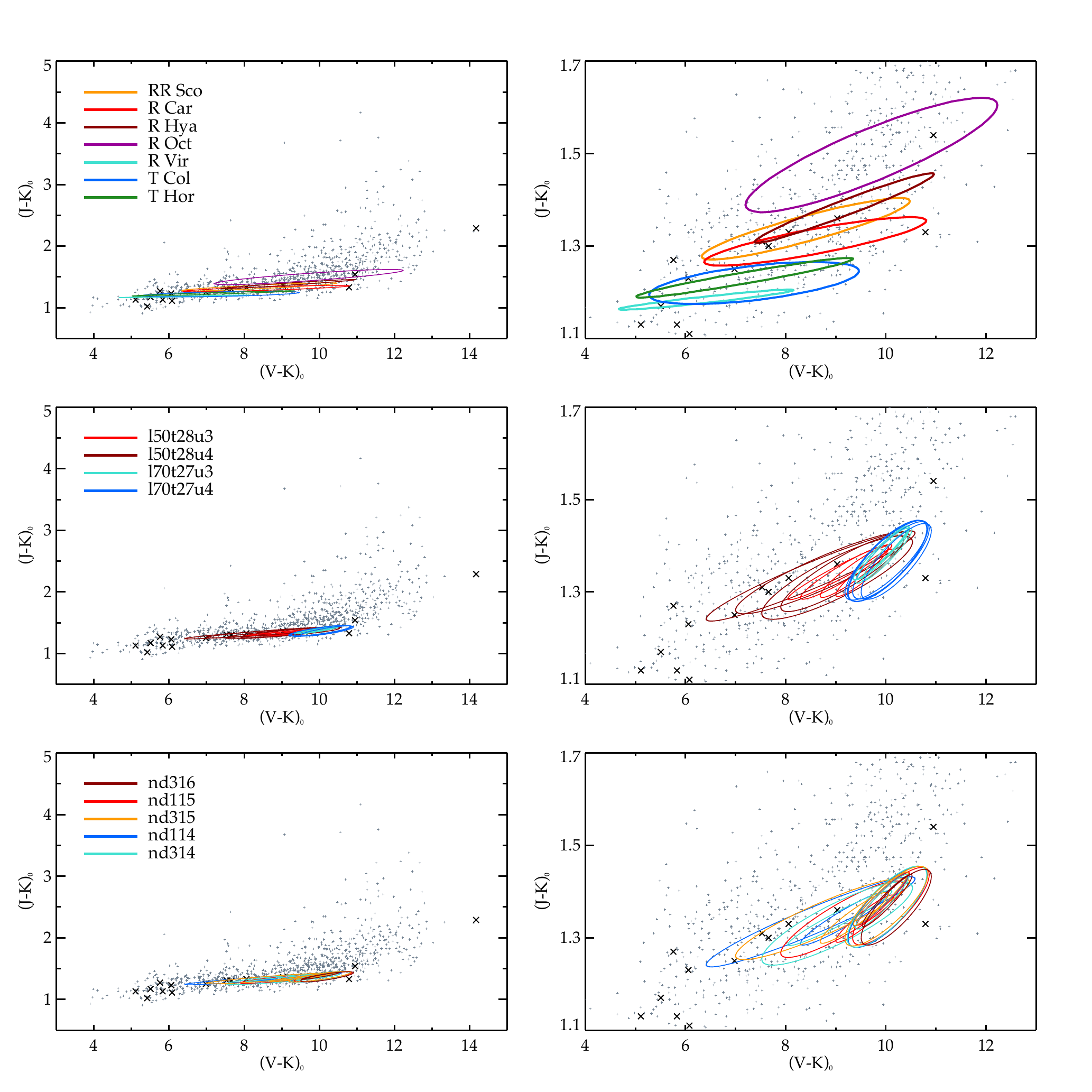}
      \caption{Observed and synthetic photometric variations of M-type AGB stars. 
{\em Top panels:} photometric variations for a sample of observed targets, derived from sine fits of light-curves; near-infrared data by \citet[][]{whitelock00} has been combined with visual data by \citet[][]{eggen75} and \citet[][]{mendoza67}; see \citet{bladh13} for details.  
{\em Middle panels:} photometric variations of the DARWIN models listed in Tab.~\ref{t_mod} with colors calculated from sine fits of the light-curves, same as for the observational data in the top panels; the lines are color-coded by stellar parameters (l50t28=A, l70t27=B) and pulsation amplitude (u3/u4). 
{\em Bottom panels:} The same model results, but color-coded by seed particle abundance (nd316 stands for $3 \cdot 10^{-16}$, nd115 for $1 \cdot 10^{-15}$, etc.). The right panels show the same content as the left panels, zoomed in and centered on the color loops. The single-epoch photometric data shown in the background of all panels represents Galactic Bulge miras \citep[][gray plus signs]{GroBlo05} and field M-type long-period variables \citep[][black crosses]{mendoza67}. 
                   }
      \label{f_phot_oli}
\end{figure*}

Since grain size is a critical factor for winds driven by photon scattering on near-transparent dust particles, any process that affects grain growth may have an impact on the mass loss mechanism. 
In our earlier papers \citep[e.g.,][]{hoefner08,bladh15} we assumed that the addition of SiO molecules from the gas is the bottleneck for the growth of the silicate grains, as in models originally developed for silicates with a significant content of Fe \citep[][]{GS99}. However, as discussed in Sect.~\ref{s_meth_ol}, for a solar element mixture the limiting factor in building Mg$_2$SiO$_4$ monomers will probably be the addition of 2~Mg atoms. The revised rates used here lead to slower grain growth, resulting in less efficient radiative acceleration. The consequences are reduced wind velocities and lower mass loss rates compared to our earlier models \citep[see values given in Tab.~\ref{t_mod} and corresponding models in][]{hoefner08}. 

A simple test of the models is to confront them with observations in a diagram showing mass loss rate vs. wind velocity. Fig.~\ref{f_dMdt_v_oli} demonstrates that the models presented in Tab.~\ref{t_mod} are consistent with observed combinations of wind properties, especially when taking into consideration that the observed mass loss rates are uncertain by up to a factor 3 \citep[for a detailed discussion see][]{ram08}. A possible exception are the models with the slowest outflows which tend to show rather high mass loss rates compared with observed low-velocity winds. Concerning the fact that all models shown in Fig.~\ref{f_dMdt_v_oli} fall into the lower left part of the diagram, we state, again, that the physical parameters of the models where chosen to produce weak to moderate winds, as such environments seem to be favorable for Al$_2$O$_3$ formation. It should, however, be noticed that higher mass loss rates and wind velocities than for model series A and B can be achieved for other combinations of parameters. \citep[see, e.g.,][]{bladh15}. 

Another important test of our models is the comparison of the resulting visual and near-IR photometry with observations. Phase-averaged synthetic (J-K) and (V-K) colors of earlier models showed good agreement with values derived from observational data \citep[see][]{bladh13,bladh15}. Even more importantly, the time-dependent behavior was found to be similar to a sample of well-observed Mira variables, i.e. flat loops in the (J-K) vs. (V-K) diagram with small variation in (J-K) and large variation in (V-K), due to variations of molecular features (in particular TiO and H$_2$O) during the pulsation cycle. This latter result was interpreted as a strong indication that the circumstellar envelopes of typical M-type AGB stars are quite transparent at visual and near-IR wavelengths, putting low upper limits on true absorption by dust. 

The new models presented here show a similar behavior,  matching observed photometric variations.
Fig.~\ref{f_phot_oli} shows the resulting loops in the (J-K) vs. (V-K) diagram, demonstrating the dependence on stellar parameters and pulsation amplitude, as well as seed particle abundance. Higher pulsation amplitude leads to larger loops, due to stronger variations of molecular abundances with phase, and the same is true for a higher effective temperature \citep[for a detailed discussion of these effects see][]{bladh13}. The abundance of seed particles, on the other hand, has a comparatively small influence on the visual and NIR colors, despite its effects on grain growth and wind properties (Tab.~\ref{t_mod}). This is consistent with our earlier results.
The outflows driven by photon scattering on large, near-transparent grains cause no significant circumstellar reddening due to absorption of stellar light by dust. The spectral energy distribution and its variation with pulsation phase is mainly influenced by the location of the condensation zone \citep[i.e, the distance from the stellar surface at which the wind-driving grains form; see][Figs.12\,\&\,14]{bladh15} which affects the structure and dynamics of the outer atmospheric layers where molecular features are formed. The revised growth rates lead to differences in the wind velocities and mass loss rates compared to our earlier models, but the distance at which Mg$_2$SiO$_4$ starts to condense is basically unchanged, resulting in similar photometric properties which are in good agreement with observations.

\subsection{Conditions for Al$_2$O$_3$ formation}\label{s_crn_cond}

Spectro-interferometric observations of circumstellar dust shells place the condensation zone of Al$_2$O$_3$ closer to the star than that of silicate grains, i.e. most likely in the pulsating atmosphere below the wind acceleration region \citep[e.g.][]{witt07,zhao11,zhao12,karo13}. On the one hand, this implies higher gas densities which favors fast grain growth. On the other hand, it means that newly-formed Al$_2$O$_3$ dust is exposed to even higher radiative flux levels than the silicates, resulting in substantial radiative heating. Since temperature is a key factor which determines both the onset of condensation and the composition of circumstellar dust we start with a detailed discussion of the optical properties of Al$_2$O$_3$ and their effects on grain temperature. Then we study the influence of other factors on Al$_2$O$_3$ condensation (i.e., stellar parameters, pulsation amplitude, seed particle abundance and resulting wind properties), and, finally, we investigate the effects of Al$_2$O$_3$ on the properties of the dynamical atmospheres and winds.

\subsubsection{Optical data for Al$_2$O$_3$}\label{s_nk_cor}

In the close vicinity of an AGB star, grain temperatures result from a balance between heating by absorption of stellar photons and cooling by thermal emission. Since the stellar photosphere is significantly hotter than the dust particles, most of the heating will typically occur at shorter wavelengths than the emission of thermal photons by the circumstellar dust. 
Consequently, reliable optical properties which cover the visual to mid-IR wavelength regions 
are of crucial importance for realistic modeling of dust formation and wind dynamics
(e.g., $n$ and $k$ data, where $m \, (\lambda) = n \, (\lambda) + i \, k \, (\lambda)$ is the complex refractive index at wavelength $\lambda$, with the imaginary part defining true absorption, and therefore radiative heating). 

In the case of Al$_2$O$_3$ the characteristic mid-IR features have been explored extensively in the astrophysical literature, both observationally and with laboratory measurements on stardust analogue materials \citep[e.g.,][]{koike95,bege97,Zeidler13}. 
The visual to near-IR absorption coefficients, on the other hand, are a major source of uncertainty. 
The reason for this is two-fold: first, Al$_2$O$_3$ is very transparent at visual and near-IR wavelengths, making measurements of $k$ technically difficult; secondly, the actual value of $k$ is strongly dependent on the microscopic structure of the material (structural defects, contamination with trace elements, etc.) which is not known {\em a priori} for dust around evolved stars. For an in-depth discussion on stardust analogues with similar general properties, in particular spinel, we refer to \citet[][]{zeidler11}. 

In order to deal with these intrinsic uncertainties in our models we take a semi-empirical approach to constrain the relevant optical properties in the visual and near-IR range for Al$_2$O$_3$ dust in AGB stars, based on the location of the condensation zone and the amount of Al condensed into grains. Using a fixed set of measured $n$ and $k$ values at mid-IR wavelengths (where measurements found in the literature differ less dramatically) we vary the value of $k$ in the visual and near-IR regime while assuming likely values for $n$ (which are much more well-determined and have little influence on the grain temperature). Comparing the location of inner edge of the condensation zone and the degree of condensation of Al resulting from our models to observations, we derive probable values and an upper limit for $k$.

\begin{figure}
\centering
\includegraphics[width=\hsize]{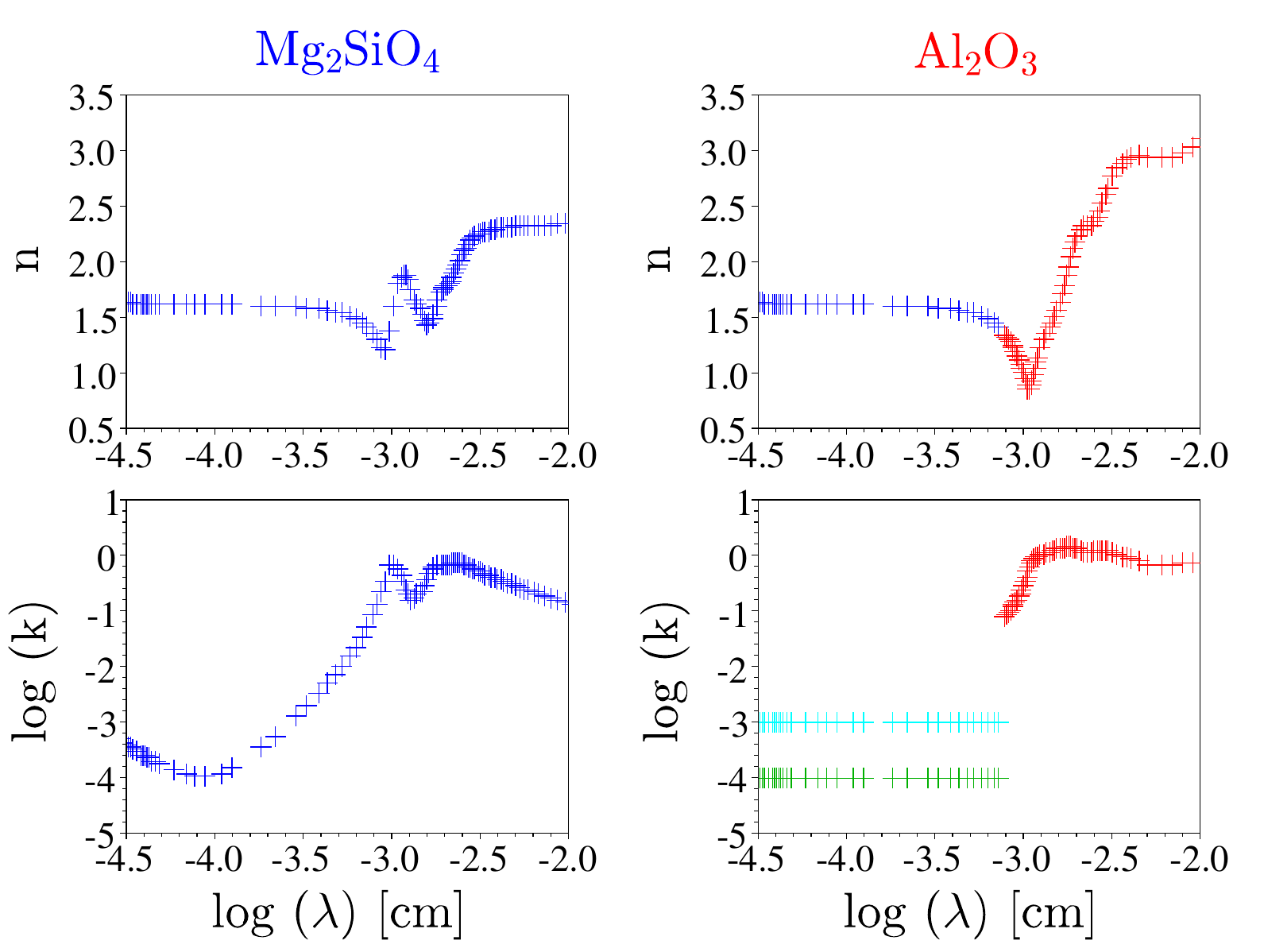}
     \caption{Refractive index data. 
                   {\em Left panels:} Mg$_2$SiO$_4$ \citep[][]{jaeger03}.   
                   {\em Right panels:} Al$_2$O$_3$ ({\em high-k} and {\em low-k} data sets, see Sec.~\ref{s_nk_cor}).
                   }
      \label{f_nk_4}
\end{figure}

\begin{figure}
\centering
\includegraphics[width=10.0cm]{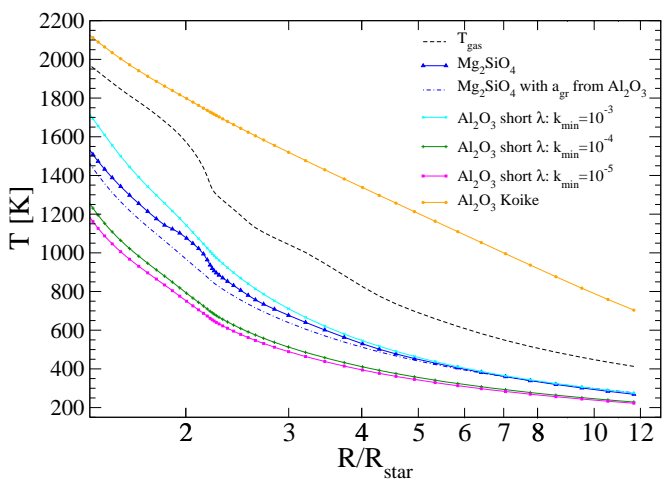}
     \caption{Radiative equilibrium temperature of dust grains as as function of distance from the star in a typical model (An315u3). The different curves correspond to different sets of optical properties (see legend and text in Sects.~\ref{s_nk_cor} and \ref{s_k_lim}).
                   }
      \label{f_t_crn}
\end{figure}

From the range of different optical properties that we have tested, we choose 2 sets, referred to as {\em high-k} and {\em low-k} data, in order to demonstrate the dependence of grain temperature and the resulting condensation fraction of Al on the uncertain values of $k$ in the visual and near-IR range. These 2 data sets for Al$_2$O$_3$ are shown in Fig.~\ref{f_nk_4}, together with the corresponding data for Mg$_2$SiO$_4$. For wavelengths longer than 7.8$\,\mu$m, the $n$ and $k$ values for Al$_2$O$_3$ are taken from \citet[][data set x2]{bege97}. For shorter wavelengths, the values of $k$ are set constant at $10^{-3}$ ({\em high-k} data set) or at $10^{-4}$ ({\em low-k} data set).\footnote{While the abrupt change to the low constant values may look dramatic, the exact form of the transition from the measured values to the constant ones has no significant effects on the results presented here, due to the relatively low radiative flux levels in this wavelength region. We have tested this by producing model with a slope similar to the silicate data shown in the left part of the figure.} 
The choice of these particular values will be motivated further below, but it should be mentioned here that they correspond to typical values measured for a material with comparable visual and near-IR properties, i.e. spinel \citep[][]{zeidler11}. The values of $n$ for wavelengths shorter than 7.8$\,\mu$m are set equal to those of Mg$_2$SiO$_4$, which seems to be a good proxy for Al$_2$O$_3$ in this regime, judging from the range of visual and near-IR data found in the literature \citep[typical values given fall between 1.5 and 1.8, with crystalline materials tending towards the upper end of this range; see, e.g.,][and references therein]{erik81,holm99}.

Figure~\ref{f_t_crn} demonstrates the influence of the optical data on the dust temperature. The {\em low-k} data leads to Al$_2$O$_3$ temperatures much lower than for Fe-free silicates. Together with the higher condensation temperature of Al$_2$O$_3$ (see Fig.~\ref{f_pstl}), this should result in significant condensation of Al$_2$O$_3$ well before silicates start forming. For the {\em high-k} data, on the other hand, the temperature of Al$_2$O$_3$ is somewhat higher than that of Fe-free silicates at the same distance from the star. Due to the significantly higher condensation temperature of Al$_2$O$_3$, however, we still expect this species to form closer to the star than silicates. As a control group we use models where $T_{\rm alu}$\,=\,$T_{\rm sil}$, i.e. the temperature of Al$_2$O$_3$ is set equal to the temperature of Mg$_2$SiO$_4$ grains of the same size at the same distance from the star (i.e. exposed to the same radiative flux). For this latter set of models we expect considerable amounts of Al$_2$O$_3$ to form before silicate condensation sets in since the condensation temperature of Al$_2$O$_3$ is about 200--300~K higher than that of Mg$_2$SiO$_4$ for typical conditions in AGB star atmospheres (see Fig.~\ref{f_pstl}).

\begin{figure}
\centering
\includegraphics[width=\hsize]{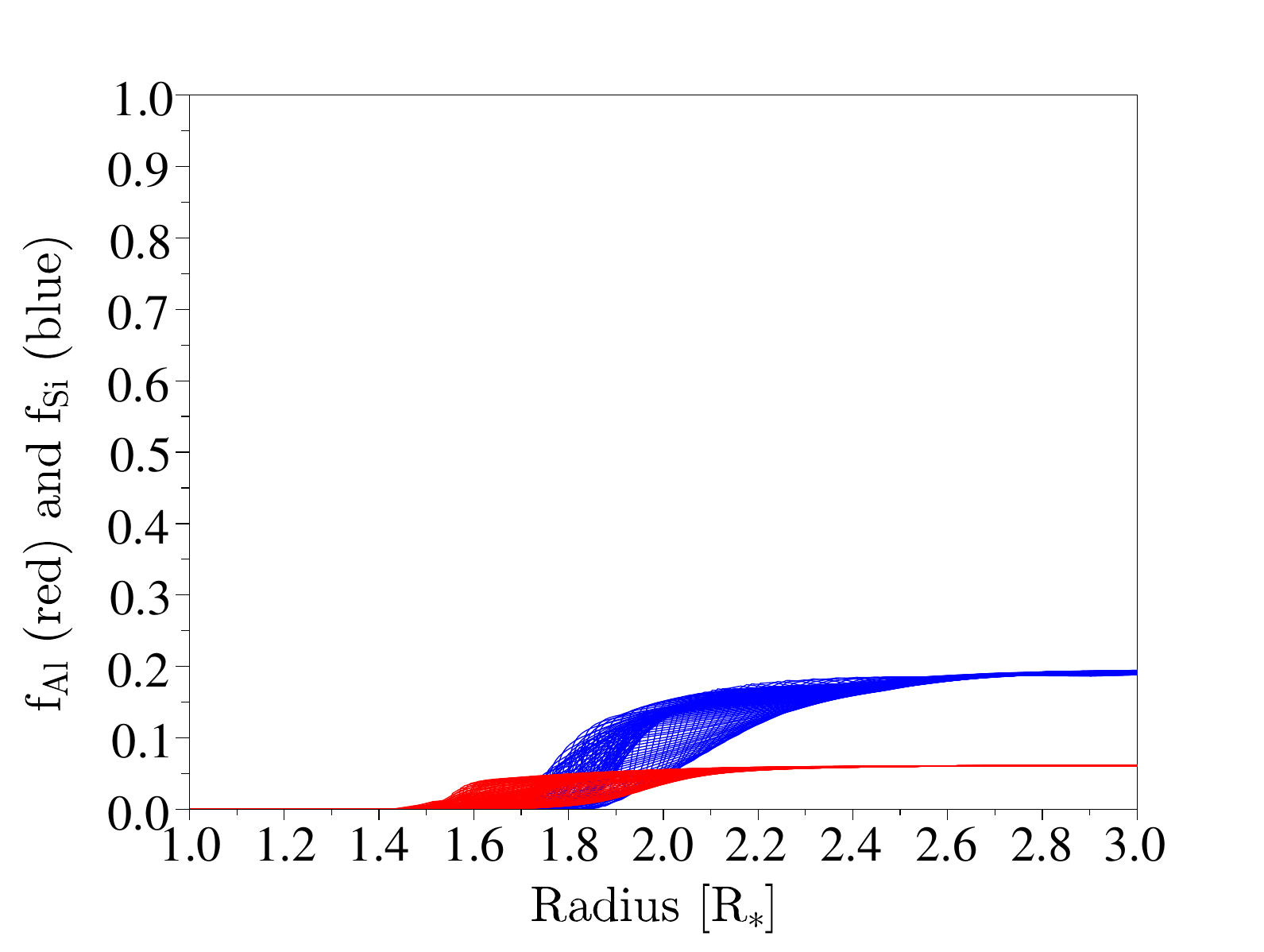}
\includegraphics[width=\hsize]{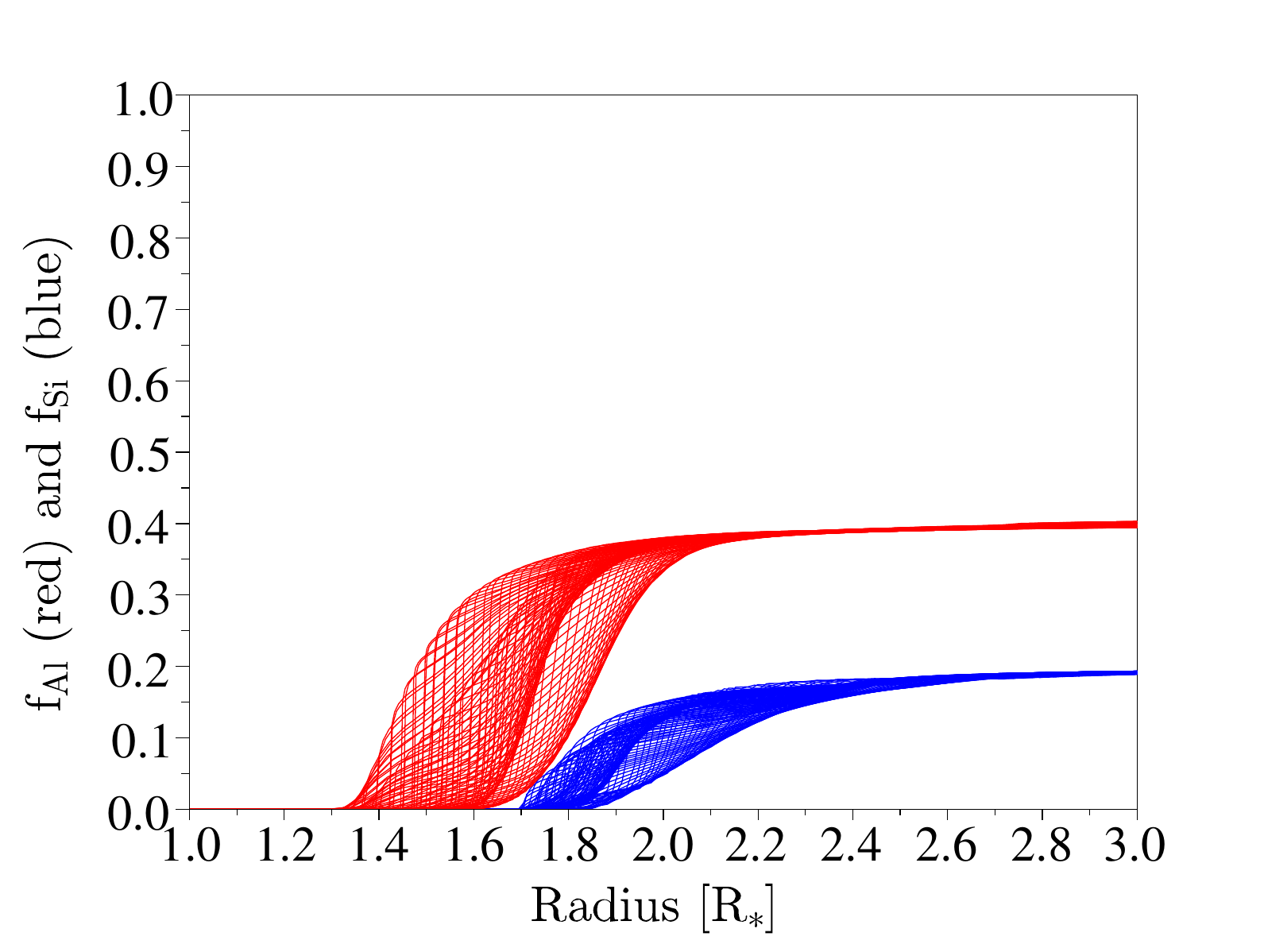}
\includegraphics[width=\hsize]{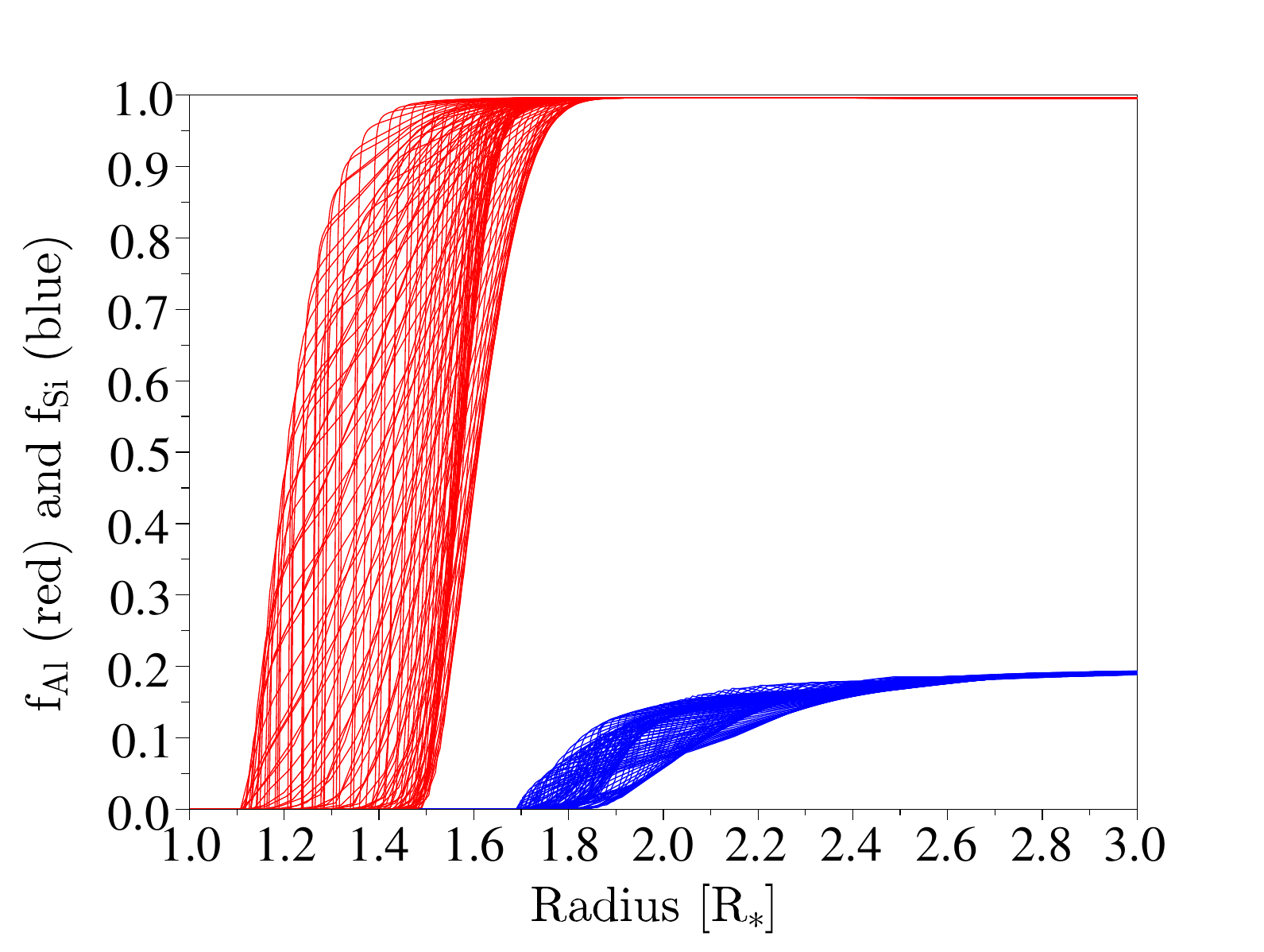}
     \caption{The time-dependent radial distribution of Al$_2$O$_3$ and silicate dust (expressed as condensation fractions of Al and Si, respectively) in model An315u3, zoomed in on the inner edge of the circumstellar dust shell (snapshots of 100 pulsation phases). The blue area (showing a similar shape in all panels) corresponds to Fe-free silicates (Mg$_2$SiO$_4$). The red curves (covering different areas in the 3 panels) represent Al$_2$O$_3$ (treated here as passive, separate grains) for different grain temperatures, resulting from the {\em high-k} optical data set (top), the {\em low-k} data (bottom), and the assumption of Al$_2$O$_3$ grain temperatures being equal to those of Mg$_2$SiO$_4$ grains of comparable sizes (middle).
                   }
      \label{f_rc_al}
\end{figure}

\subsubsection{Effects of optical properties on Al$_2$O$_3$ formation}

Since the influence of optical properties on the formation of Al$_2$O$_3$ is most easily seen against the backdrop of given atmosphere and wind structures, we start with models where Al$_2$O$_3$ is treated as a passive dust species, i.e. solving the equations of grain growth but ignoring potential effects of Al$_2$O$_3$ on dynamics or other models properties (e.g., its contribution to radiation pressure, etc.). The resulting fractions of Al condensed into grains are given in Tab.~\ref{t_mod} for 3 cases, i.e., {\em high-k} data, {\em low-k} data, and assuming that the temperature of Al$_2$O$_3$ is similar to Fe-free silicate grains of equal size. 

Models computed with the {\em high-k} data set (corresponding to the strongest radiative heating and, consequently, the highest grain temperatures of the 3 cases investigated here, see Fig.~\ref{f_t_crn}) span a wide range of Al condensation fractions, from a few percent to complete condensation, depending on the other model parameters which affect the efficiency of grain growth. In contrast, the models based on the {\em low-k} data set (corresponding to the lowest of the three Al$_2$O$_3$ temperatures at a given pulsation phase and distance from the star) show complete condensation for all but 2 models, i.e. those of series A with the lowest seed particle abundances (even they have only a few percent of Al left in the gas phase). The control group with $T_{\rm alu}$ = $T_{\rm sil}$ falls somewhere between the {\em high-k} and {\em low-k} cases, as expected from the grain temperatures. Clearly, for a given combination of stellar parameters, pulsation amplitude and seed particle abundance, the temperature of the Al$_2$O$_3$ grains determines the fraction of Al condensed into grains, with lower temperatures leading to an earlier onset of Al$_2$O$_3$ condensation and therefore higher $f_{\rm Al}$ values. 

This effect is illustrated in Fig.~\ref{f_rc_al} for a selected model: The panels show the condensation fraction of Si and of Al for the case of $T_{\rm alu}$ = $T_{\rm sil}$ (middle), as well as for the cases of {\em high-k} (top) and {\em low-k} (bottom). When assuming temperatures for Al$_2$O$_3$ similar to Fe-free silicates exposed to the same radiative flux, Al condensation starts earlier (closer to the star) than Si condensation, due to the higher thermal stability of Al$_2$O$_3$ (Fig.~\ref{f_rc_al}, middle). However, as soon as the silicate grains reach sizes resulting in significant radiative acceleration, the further growth of the Al$_2$O$_3$ grains is quenched by falling densities in the outflow, leading to a condensation fraction of Al well below 100\,\%. This effect is even more pronounced for the {\em high-k} case (Fig.~\ref{f_rc_al}, top). Since the Al$_2$O$_3$ temperature is higher for a given pulsation phase and distance from the star than in the previous case, the inner edge of the condensation zone is located further from the stellar surface, allowing for less grain growth before silicate condensation and wind acceleration quench Al$_2$O$_3$ condensation at a smaller $f_{\rm Al}$ value. The lower Al$_2$O$_3$ temperatures resulting from the {\em low-k} data set, on the other hand, lead to the opposite effect, allowing for full condensation of Al before wind acceleration sets in (Fig.~\ref{f_rc_al}, bottom). 
It should be noted in this context that differences in $f_{\rm Al}$ values resulting from the chosen optical properties may be more or less pronounced than for the model shown in Fig.~\ref{f_rc_al}, depending on the overall efficiency of dust formation (which is also affected by the stellar parameters, pulsation properties and seed particle abundances, as will be discussed below). The qualitative trends are, however, similar for other combinations of parameters.

\subsubsection{Optical properties: limiting cases}\label{s_k_lim}

At this point the question arises if lowering the visual and near-IR value of $k$ below $10^{-4}$ will lead to even more favorable conditions for Al$_2$O$_3$ formation through lower grain temperatures than in the {\em low-k} case. A simple test, however, shows that this is not very likely. While going from $10^{-3}$ ({\em high-k} case) to $10^{-4}$ ({\em low-k} case) leads to a reduction of the Al$_2$O$_3$ temperature in the dust formation zone of about 400\,K, a further reduction to $10^{-5}$ only lowers the grain temperature by about 50\,K  (see Fig.~\ref{f_t_crn}). The corresponding absorption coefficients in the visual and near-IR regime have obviously reached a level where the heating of the grains is no longer dominated by the stellar flux in this wavelength region. Instead, the grain temperature is set by a balance of absorption and emission at longer wavelengths where the opacities are substantially higher. The results obtained with the {\em low-k} data set should therefore be representative of grains with very high transparency in the visual and near-IR regime, in general. For comparison, a model by \citet[][with $\Tstar = 2500\,$K and $\Lstar = 10000\,\Lsun$]{woit06b}, using absorption coefficients for Al$_2$O$_3$ in the visual and near-IR regime which are more than an order of magnitude lower than those of Mg$_2$SiO$_4$, shows  substantial amounts of Al$_2$O$_3$ around $1.5\,\Rstar$, which is in good agreement with the {\em low-k} case (see Fig.~\ref{f_rc_al}). 

Raising the $k$ values above $10^{-3}$, on the other hand, does not lead to a similar saturation effect. On the contrary, the grain temperature keeps rising significantly with the visual and near-IR values of $k$ (due to increased radiative heating by absorption), and, consequently, the condensation zone moves further away from the star. An extreme example is given in Fig.~\ref{f_t_crn}, i.e. the temperatures resulting from the opacity data of \citet[][data set ISAS, corresponding to visual and near-IR $k$ values of about $10^{-1}$ to $10^{-2}$]{koike95}. 
In this case the temperature does not drop below the condensation temperature of Al$_2$O$_3$ (about 1400\,K, see Fig.~\ref{f_pstl}) until far beyond the zone where silicates start to form (i.e. at distances of about 4 and 2 stellar radii, respectively, in this example). Under these circumstances Al$_2$O$_3$ condensation will be suppressed efficiently. Considering that the further growth of silicates is quenched by falling densities as soon as radiative acceleration drives the gas away from the star, Al$_2$O$_3$ condensation should not occur beyond this point. The growth rates of Al$_2$O$_3$ will be substantially lower than those of silicates, which is mainly due to the much lower elemental abundance of Al compared to Si and Mg (see Eqs.\,(\ref{e_jgr_crn}) and (\ref{e_jgr_ol}), respectively; note that the thermal velocities of Al and Mg are comparable). Test models using the data of \citet[][]{koike95} instead of the {\em low-k} or {\em high-k} data are, indeed, not producing Al$_2$O$_3$ dust. In other words, the fact that high quantities of Al$_2$O$_3$ are observed in AGB stars, and at closer distances than silicates, implies a true upper limit for the visual and near-IR values of $k$, well below those given by \citet[][]{koike95}.

\subsubsection{Influence of other parameters}

After demonstrating the effects of the optical properties, we now turn to the influence of other quantities on Al$_2$O$_3$ condensation. The following discussion refers mainly to the models based on the {\em high-k} data set and the control group assuming $T_{\rm alu}$ = $T_{\rm sil}$ which show similar trends of the $f_{\rm Al}$ values with various model parameters. These trends are not (or only weakly) apparent in the {\em low-k} models since they all have Al condensation close to 100\,$\%$. 

For fixed stellar parameters (series A or B) and fixed amplitude of pulsation (here $\Delta u_{\rm P}$ = 3 or 4 km/s), the fraction of Al condensed into grains tends to increase with an increasing abundance of seed particles $n_d/n_{\rm H}$ (see Tab.~\ref{t_mod}). This is easy to understand, since more seed particles correspond to a larger total grain surface area in a given volume of gas, and therefore a higher rate of consumption of Al by grain growth. The same is true for silicates (see $f_{\rm Si}$ values given in Tab.~\ref{t_mod}). 

A higher pulsation amplitude (for all other input parameters being equal) also results in more Al$_2$O$_3$ dust (i.e. higher $f_{\rm Al}$) which is due to stronger compression of the gas in shocks and the resulting effects on growth rates. In this respect, however, silicate condensation behaves differently, leading to comparable values of  $f_{\rm Si}$ for both pulsation amplitudes, due to a self-regulating effect in grain growth for the wind-driving dust species (i.e., grain growth quickly coming to halt due to the falling densities in the outflow, triggered when the grains have reached a size which produces sufficient radiative acceleration).

Finally, a comparison of individual models in series A and B, for equal pulsation amplitudes (i.e. velocity amplitudes at the inner boundary) and seed particle abundances, demonstrates the influence of the stellar parameters. The higher effective temperature of the A-models leads to systematically lower amounts of Al$_2$O$_3$ dust, compared with the corresponding B-models, for all but the {\em low-k} cases which show (almost) complete condensation of Al. The lower effective temperature and higher luminosity of the B-models are favorable for both grain growth (condensation closer to the stellar photosphere, at higher densities) and radiative acceleration, leading to higher mass loss rates and wind velocities than in the corresponding A-models, despite somewhat lower silicate condensation.

\subsection{Effects of Al$_2$O$_3$ on atmosphere \& wind dynamics}\label{s_crn_eff}

\begin{table}
\caption{\label{t_mod_co} Parameters and resulting dust properties of {\em Al$_2$O$_3$ only} 
models. Note that none of the models develops a wind.}
\centering
\begin{tabular}{l|cc|cc}
\hline\hline
  & & & &  \\
  model &  $n_d/n_{\rm H}$  &  $\Delta u_{\rm P}$ &  $a_{gr}$  &  $f_{\rm Al}$  \\
  name  &                              & [km/s]                    & [$\mu$m]  & $T_{\rm alu}$\,=\,$T_{\rm sil}$ \\
  & & & &  \\ 
\hline
  & & & &  \\ 
  An315u3ao & $3.0 \cdot 10^{-15}$ & 3.0 & 0.12 & 0.3 \\ 
  An114u3ao & $1.0 \cdot 10^{-14}$ & 3.0 & 0.09 & 0.5 \\ 
  & & & & \\
  An315u4ao & $3.0 \cdot 10^{-15}$ & 4.0 & 0.16 & 0.8 \\ 
  An114u4ao & $1.0 \cdot 10^{-14}$ & 4.0 & 0.11 & 0.9 \\ 
  & & & &  \\ 
  Bn316u3ao & $3.0 \cdot 10^{-16}$ & 3.0 & 0.37 & 0.97 \\ 
  Bn115u3ao & $1.0 \cdot 10^{-15}$ & 3.0 & 0.25 & 1.00 \\ 
  & & & & \\
\hline
\end{tabular}
\end{table}

So far, we have studied which conditions are necessary to allow for the formation of Al$_2$O$_3$ at the close distances and in the high quantities indicated by observations. For simplicity, we have been treating Al$_2$O$_3$ as a passive dust species, ignoring its potential influences on the atmosphere and wind (e.g. neglecting its radiation pressure). In other words, we have investigated how Al$_2$O$_3$ is affected by its surroundings, but not what effects it may have on them, in turn. In the literature, Al$_2$O$_3$ has been discussed as a potential contributor to wind driving and as a seed particle for the growth of silicate grains \citep[e.g.][]{KoSo97a,KoSo97b}. In the following, we use our models to shed new light on these suggestions. In order to keep the discussion focused, we will use only one of the three options for the grain temperature discussed above, i.e. the case of the Al$_2$O$_3$ temperature being equal to that of Fe-free silicate grains with the same size. Such models represent some middle ground between the {\em high-k} and {\em low-k} cases while still allowing for considerable condensation fractions of Al for typical stellar parameters.

\subsubsection{Radiation pressure on Al$_2$O$_3$}

We start with the question if Al$_2$O$_3$ will contribute directly to wind driving, i.e. through radiation pressure. From our experiments with the visual and NIR optical data, we concluded that $k$ is probably less than $10^{-3}$ in this wavelength region, i.e comparable to Fe-free silicates which are notoriously transparent. Considering, in addition, the lower abundance of Al than Si and Mg, the radiation pressure resulting from true absorption by Al$_2$O$_3$ can be regarded as negligible. As for the Fe-free silicates, any noticeable radiative acceleration would have to be due to photon scattering by grains with sizes of about 0.1 -- 1 microns. Measurements of $n$ indicate similar visual and NIR values for the two materials, implying similar scattering efficiency for grains of comparable sizes. 

Based on this estimate, we can compute the relative value of Al$_2$O$_3$ and silicate opacities contributing to radiative acceleration, using the equations and numbers given in Sect.~\ref{s_methods}. Assuming equal grain radii and efficiency factors $Q_{\rm acc}$, we find
\begin{equation}
   \frac{\kappa_{\rm acc}^{\rm alu}}{\kappa_{\rm acc}^{\rm sil}}  \approx 0.3 \, \frac{\varepsilon_{\rm Al}}{\varepsilon_{\rm Si}} \, \frac{f_{\rm Al}}{f_{\rm Si}}  \approx 0.025 \, \frac{f_{\rm Al}}{f_{\rm Si}} \,\, ,
\end{equation}
with Al/Si being about 8\% by number in a solar mixture. This indicates that  $f_{\rm Al}$, i.e. the condensation fraction of Al, needs to be about a factor of 40 higher than that of Si, in order to produce a value of opacity (and consequently radiative acceleration) which is comparable to the wind-driving Fe-free silicate grains.
Regarding the condensation fractions of Si necessary to drive winds for the models listed in Tab.~\ref{t_mod}, we note that this corresponds to more material than is available for building Al$_2$O$_3$ (with full condensation of Al corresponding to $f_{\rm Al} = 1$). Considering that the lowest condensation fractions of Si are 0.17 for model series A and 0.13 for model series B, an increase in the abundance of Al by at least a factor of 7 and 5, respectively, would be required in order to reach $\kappa_{\rm acc}^{\rm alu} / \kappa_{\rm acc}^{\rm sil} = 1$.
Therefore, we do not expect Al$_2$O$_3$ to be a wind driver for typical AGB stars. 

Another way to test the possible effects of radiation pressure by Al$_2$O$_3$ is to produce dynamical models which take this force into account, but where silicate formation is switched off. Any resulting outflow would then be due to Al$_2$O$_3$. This approach has the added benefit, compared to the simple estimate given above, that it also allows to account for the influence of pulsation-induced shocks and the variable luminosity in a more consistent way. The corresponding models are referred to as {\em {Al$_2$O$_3$} only} in the following discussion, and their properties are given in Tab.\ref{t_mod_co} (labeled with {\em ao} in the model names). As expected, none of these {\em {Al$_2$O$_3$} only} models develops a wind, despite considerable condensation fractions of Al (reaching up to 100\% for the cooler, more luminous models of series B). In fact, the resulting pulsating atmospheres show mean extensions which are quite similar to dust-free models, in accordance with the low levels of radiative pressure on Al$_2$O$_3$ estimated above.

\begin{table*}
\caption{\label{t_mod_cm}Model parameters and resulting wind properties of  {\em core-mantle grain} models. See Tab.~\ref{t_mod} and text for a definition of symbols. 
}
\centering
\begin{tabular}{l|cc|ccccccc}
\hline\hline
  & & & & & & & \\ 
  model &  $n_d/n_{\rm H}$  &  $\Delta u_{\rm P}$ &  $\dot{M}$  & $u_{\rm ext}$  & $f_{\rm Si}$ & $f_{\rm Mg}$  &  $f_{\rm Al}$ & $a_{\rm core}$ & $a_{\rm gr}$   \\
  name  &  & [km/s] & [$\Msun$/yr] & [km/s] & & & $T_{\rm alu}$\,=\,$T_{\rm sil}$ & [$\mu$m] & [$\mu$m]  \\
  & & & & & & & \\ 
\hline
  & & & & & & & \\ 
  An315u3cmg & $3.0 \cdot 10^{-15}$ & 3.0 & $5 \cdot 10^{-7}$ & 7 & 0.19 & 0.36 & 0.54 & 0.14 & 0.35 \\ 
  An114u3cmg & $1.0 \cdot 10^{-14}$ & 3.0 & $7 \cdot 10^{-7}$ & 9 & 0.24 & 0.45 & 0.76 & 0.11 & 0.25 \\ 
  & & & & & & & \\ 
  An315u4cmg & $3.0 \cdot 10^{-15}$ & 4.0 & $6 \cdot 10^{-7}$ & 7 & 0.17 & 0.32 & 0.82 & 0.16 & 0.34\\ 
  An114u4cmg & $1.0 \cdot 10^{-14}$ & 4.0 & $7 \cdot 10^{-7}$ & 11 & 0.26 & 0.49 & 0.75 & 0.10 & 0.26 \\ 
  & & & & & & & \\ 
  Bn316u3cmg & $3.0 \cdot 10^{-16}$ & 3.0 & $8 \cdot 10^{-7}$ & 4 & 0.11 & 0.21 & 0.75 & 0.34 & 0.64 \\ 
  Bn115u3cmg & $1.0 \cdot 10^{-15}$ & 3.0 & $2 \cdot 10^{-6}$ & 8 & 0.13 & 0.24 & 0.50 & 0.20 & 0.44 \\ 
  & & & & & & & \\ 
\hline
\end{tabular}
\end{table*}

\subsubsection{Composite grains: Al$_2$O$_3$ core, silicate mantle}

Even if Al$_2$O$_3$ dust is not likely to play a direct role for wind driving through radiative pressure, it may affect the mass loss process in more subtle, indirect ways. A topic which has been discussed in the literature \citep[e.g.][]{KoSo97a,KoSo97b} is if Al$_2$O$_3$ grains could be seed particles for the subsequent condensation of silicate dust, further out in the atmosphere at lower temperatures. A possible consequence of this scenario is a speed-up of grain growth to sizes relevant for driving a wind by scattering of photons. When building silicate mantles on fully-grown Al$_2$O$_3$ grains, it may take considerably less time to reach the critical regime of grain sizes, counting from the onset of silicate condensation. The effect is similar to an increase of the growth rate for pure silicate grains starting from tiny seed particles. As discussed in Sect.~\ref{s_rates}, higher growth rates tend to produce faster winds and higher mass loss rates. Such changes of the wind properties, however, may in turn influence the conditions in the deeper layers of the atmosphere where the Al$_2$O$_3$ particles are formed, potentially leading to an intricate feedback. 

To study these processes, we have computed dynamical models where Mg$_2$SiO$_4$ mantles may condense on top of Al$_2$O$_3$ cores, taking into account both, the time-dependent growth of each species, and the total scattering efficiency of the composite grains which is relevant for the radiative pressure. In the following discussion, theses models are referred to as {\em core-mantle grain} models and their names are labeled with {\em cmg}. Since observations show more pronounced features of Al$_2$O$_3$ in objects with lower mass loss rates, we focus on models from series A to demonstrate the effects of core-mantle grains on wind properties. In addition, a few B-models are used to confirm trends for a wider range of parameters (see Tab.~\ref{t_mod_cm}). 

When comparing the results for individual {\em core-mantle grain} models to the corresponding models driven by pure silicate grains (with Al$_2$O$_3$ treated as a passive separate species,  Tab.~\ref{t_mod}), we see that the {\em core-mantle grain} models show faster winds with higher mass loss rates, as expected from the arguments given above (note in this context that the radii of the Al$_2$O$_3$ cores are a substantial fraction of the total grain radius). The total sizes of the composite grains, however, are strikingly similar to the wind-driving pure silicate grains in the original models. This clearly demonstrates the self-regulating feedback between wind acceleration and dust condensation discussed before, i.e. efficient quenching of further grain growth in the wind acceleration zone due to falling densities.  

It should be noted that this self-regulating mechanism due to radiative acceleration applies to the total size of the composite grains, and therefore mostly to the silicate mantle. The Al$_2$O$_3$ grains which eventually become the cores of the composite particles are not affected by a similar mechanism, due to their much lower radiation pressure. On the other hand, the efficiency of Al$_2$O$_3$ condensation depends on the conditions in the atmospheric layers where this species is formed, which in turn may be influenced by the dynamics of the wind acceleration zone. As we have seen from the experimental {\em {Al$_2$O$_3$} only} models, even pulsating atmospheres without a wind can lead to considerable fractions of Al condensation (cf.~Tab.~\ref{t_mod_co}). Intuitively, one might expect that an outflow originating above the Al$_2$O$_3$ condensation zone (e.g. driven by silicate dust) may enhance Al$_2$O$_3$ formation by causing higher densities in the outer atmospheric layers. 

To test this idea, we compare the condensation fractions of Al in the windless {\em {Al$_2$O$_3$} only} models with the wind-forming {\em core-mantle grain} models (see Tab.~\ref{t_mod_co} and Tab.~\ref{t_mod_cm}, respectively). We start with two pairs of models, An315u3ao/cmg and An114u3ao/cmg, which are characteristic of a group with moderate pulsation amplitudes, mass loss rates and wind velocities. In this case, the wind-forming {\em cmg} models show, indeed, significantly higher values of $f_{\rm Al}$ than their wind-less {\em ao} counterparts. 
Increasing the pulsation amplitude while keeping all other parameters constant leads to models An315u4ao/cmg and An114u4ao/cmg (second group in Tab.~\ref{t_mod_co} and Tab.~\ref{t_mod_cm}), with somewhat higher mass loss rate (An315u4cmg) or wind velocity (An114u4cmg). In contrast to the first group of models with the lower pulsation amplitude, the condensation fraction of Al is similar for An315u4ao/cmg and actually lower in the wind-forming model An114u4cmg than in its wind-less counterpart An114u4ao, reversing the expected trend. 

The probable cause is that we have based our argument on atmospheric densities alone. A faster transport of material through the condensation zone will reduce the time available for grain growth. Winds can therefore have both positive and negative effects on Al$_2$O$_3$ condensation.  
To demonstrate this even more clearly, we add some models from series B to our sample. Due to the higher luminosity, wind acceleration is more efficient than for the A-models, and both wind-forming B-models, i.e. Bn316u3cmg and Bn115u3cmg, show significantly less condensation of Al$_2$O$_3$ than their wind-less counterparts, Bn316u3ao and Bn115u3ao, respectively. Finally, it should be mentioned here that the $f_{\rm Al}$ values of the original wind models (driven by pure silicate grains, see Tab.~\ref{t_mod}) fall generally between the values of the {\em ao} and {\em cmg} models, in accordance with the lower mass loss rates and wind velocities compared to the {\em cmg} models.

In summary, our wind-forming {\em core-mantle grain} models show that Al$_2$O$_3$ tends to have a favorable effect on mass loss rates and wind velocities while the opposite is not necessarily true: A comparison with windless {\em {Al$_2$O$_3$} only} models indicates that a moderate stellar wind can be beneficial for Al$_2$O$_3$ condensation, but that this trend may be reversed for strong outflows.

\begin{figure*}
\centering
\includegraphics[width=13cm,angle=270]{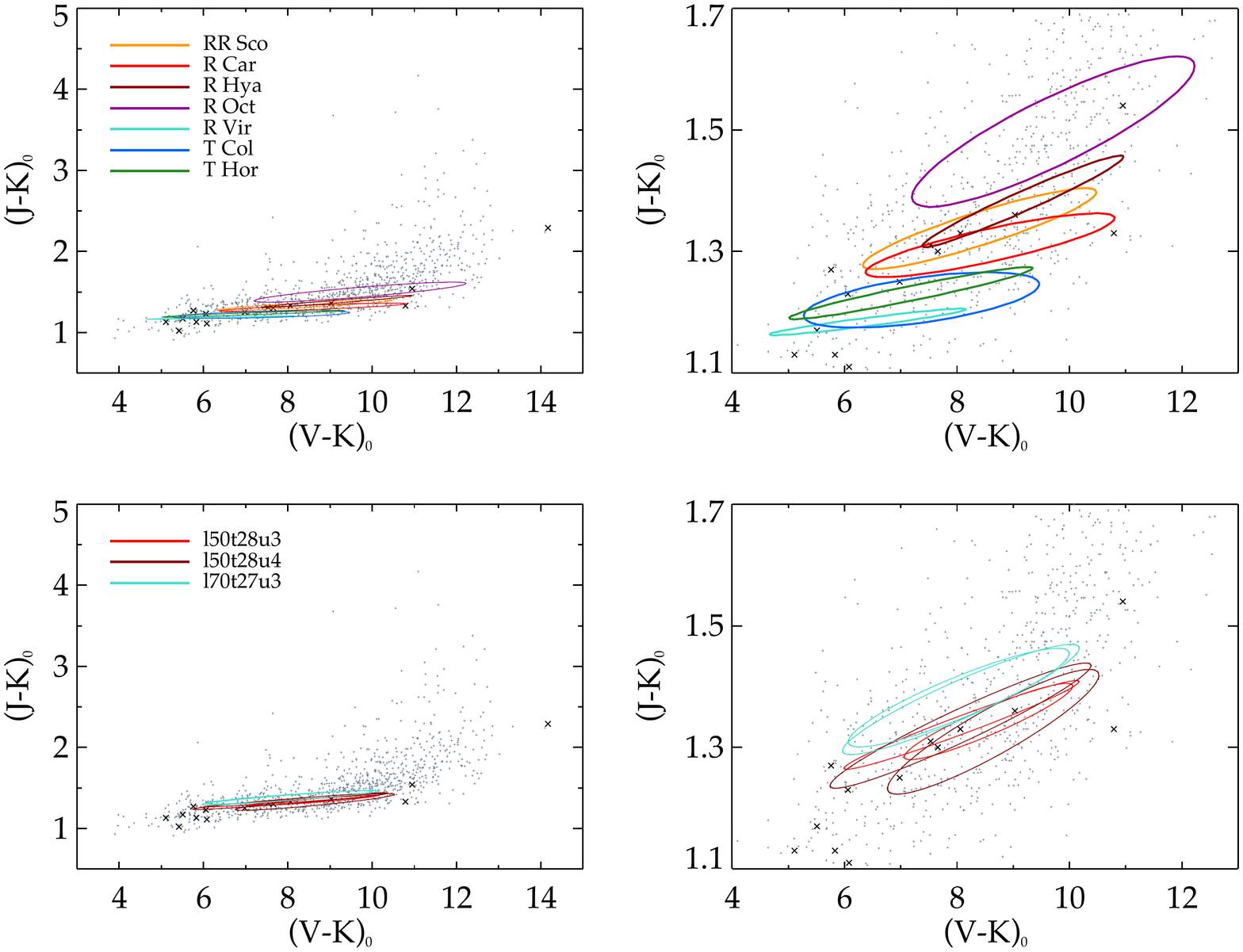}
      \caption{Observed and synthetic photometric variations of M-type AGB stars. {\em Upper panels:} photometric variations for a sample of observed targets, derived from sine fits of light-curves (see Fig.~\ref{f_phot_oli} and related text for details). {\em Lower panels:} photometric variations for the {\em core-mantle grain} models listed in Tab.~\ref{t_mod_cm} with colors calculated from sine fits of the light-curves, same as for the observational data in the top panels; the lines are color-coded by stellar parameters and pulsation amplitude. 
                   }
      \label{f_phot_cmg}
\end{figure*}

Finally, after discussing the effects of core-mantle grains on mass loss and Al condensation fractions, we turn to the visual \& near-IR photometric properties of the {\em core-mantle grain} models. Fig.~\ref{f_phot_cmg} shows the variation of (J-K) vs. (V-K) colours during a pulsation cycle. Compared to the wind models driven by pure silicate grains (Fig.~\ref{f_phot_oli}) the {\em cmg} models show a larger range of variation in (V-K), bringing the loops into almost perfect agreement with observations. Remarkably, the loops resulting from all 6 {\em cmg} models are rather similar in size and shape, in contrast to the wind models based on pure silicate grains. As discussed by \citet[][]{bladh13}, the variation in (V-K) is due to changes in molecular features (in particular TiO which dominates the V band). Consequently, the larger variations in (V-K) shown by the {\em cmg} models are an indication of stronger temporal variations in the molecular layers, below the wind acceleration region. As discussed above, the direct dynamical effects of Al$_2$O$_3$ (which forms closer to the star than the Fe-free silicates that trigger the outflow) are rather insignificant, so the stronger variations are probably due to indirect effects of the more efficient wind acceleration in the {\em cmg} models (caused by a speed-up of grain growth). It should also be noted in this context that the combinations of wind velocities and mass loss rates for the {\em core-mantle grain} models -- while different from those of winds driven by pure silicate grains -- are still in agreement with observed values.

\begin{figure*}
\centering
\includegraphics[width=\hsize]{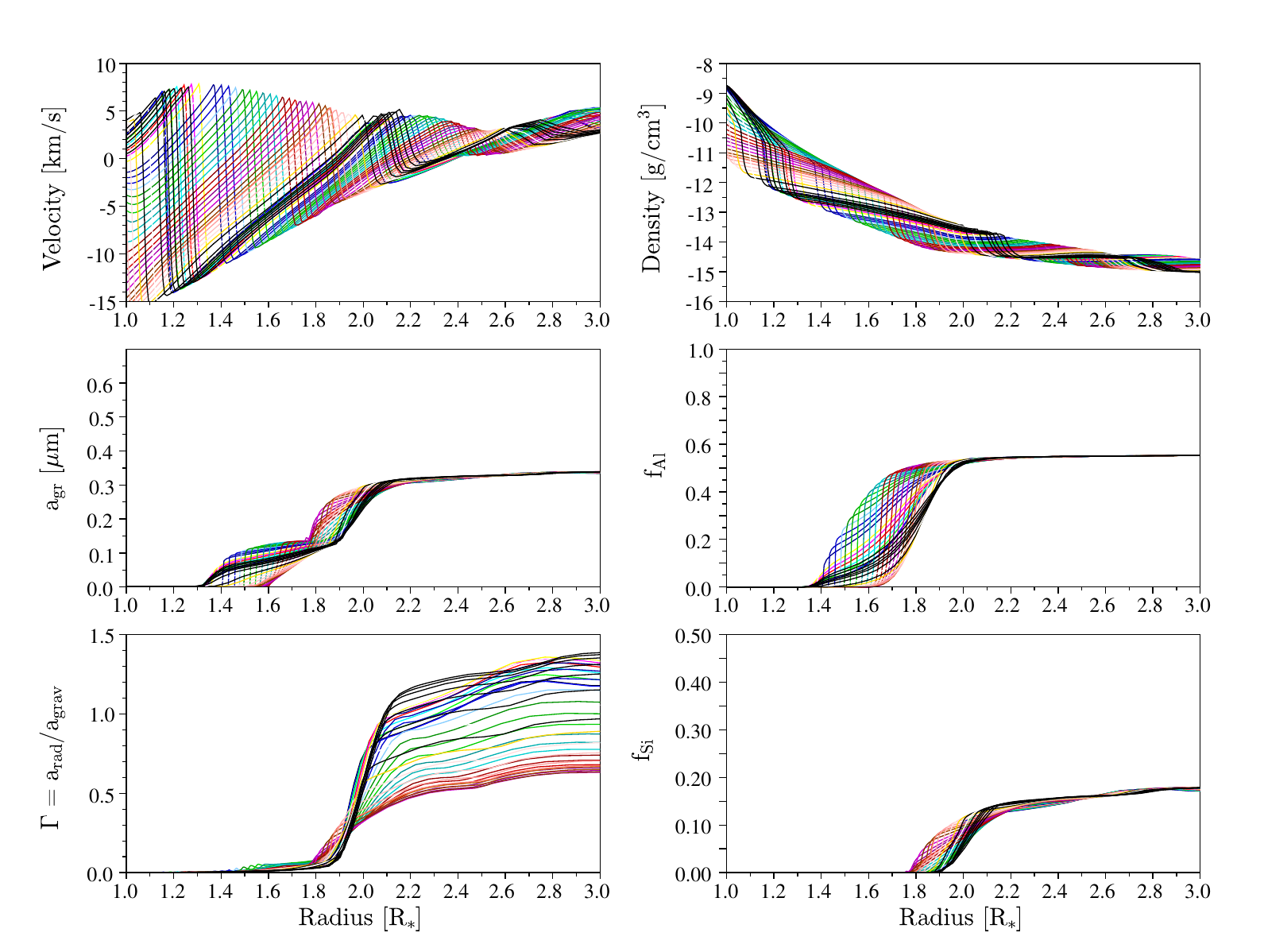}
     \caption{The time-dependent radial structure of model An315u3cmg, zoomed in on the dust formation region (snapshots of 40 pulsation phases). {\em Left, top to bottom}: Velocity, the radius of the composite grains and the ratio of radiative to gravitational acceleration; {\em Right, top to bottom}: Gas density, Al condensation fraction, Si condensation fraction.                   }
      \label{f_cmg}
\end{figure*}

\section{Discussion: Model results and observations}\label{s_discussion}

In Sect.~\ref{s_results} we have presented 3 sets of DARWIN models with different treatments of Al$_2$O$_3$ dust (i.e., as a passive species in addition to wind-driving silicates, as the only dust species contributing to radiative pressure, and as grains consisting of an Al$_2$O$_3$ core and a silicate mantle). Each type of model was constructed to answer specific questions about the necessary conditions for Al$_2$O$_3$ formation and the effects of Al$_2$O$_3$ on the structure and dynamics of atmospheres and winds. Now we will discuss in more detail how these results relate to observations and laboratory measurements. We start with the conditions in the Al$_2$O$_3$ condensation zone and then proceed to feedback mechanisms between Al$_2$O$_3$, silicates and wind properties.

\subsection{Visual and near-IR optical properties of Al$_2$O$_3$}

Spectro-interferometric studies of circumstellar dust shells at high angular resolution indicate high concentrations of Al$_2$O$_3$ at distances of about 2 stellar radii, or less \citep[e.g.][]{witt07,zhao11,zhao12,karo13}. This close to the star, the grains will be subject to substantial heating by the strong radiative flux from the stellar photosphere. Therefore their very existence (which implies grain temperatures below the condensation temperature) puts constraints on their absorption properties at visual and near-IR wavelengths. In order to obtain significant amounts of Al$_2$O$_3$ for typical stellar parameters in our dynamical models, we have to assume that the imaginary part of the refractive index, i.e. $k$ (which defines true absorption, and therefore radiative heating), is about $10^{-3}$ or less at visual and near-IR wavelengths where the stellar flux peaks. 

This upper limit is much lower (about 1-2 orders of magnitude) than the values given by \citet{koike95} for Al$_2$O$_3$, but comparable to measurement of $k$ for a natural sample of spinel by \citet{zeidler11}. Both Al$_2$O$_3$ and spinel (MgAl$_2$O$_4$) in their pure forms are very transparent in the visual and near-IR regime (which makes them interesting materials for technical applications due to their high mechanical stability). The still quite low but measurable values of $k$ for the spinel sample studied by \citet{zeidler11} are due to impurities of Cr and Fe, about 1\,\% of each, i.e. levels that are compatible with the relative abundances of Al and Cr in a solar mixture. It seems therefore plausible that circumstellar Al$_2$O$_3$ has similar $k$ values in the visual and near-IR due to impurities. In this context it is interesting to note that gem-quality Al$_2$O$_3$ containing small amounts of Cr is better known as ruby. This means that AGB stars could be a source of sub-micron-sized cosmic rubies.

\subsection{Gravitationally bound dust shells}

From a theoretical point of view, there are several reasons why the observed high concentrations of dust in the close vicinity of AGB stars are interesting. Apart from the constraints on the optical properties, as discussed above, these observations give insights into the structure and dynamics of atmospheres and wind acceleration zones. As demonstrated in several papers \citep[e.g.][]{ireland05,karo13,khouri15}, a simple extrapolation of steady wind structures inwards to the regions where the inner edges of the dust shells are located leads to contradictions. In particular, the large amounts of Al$_2$O$_3$ implied by mid-IR spectra would require Al abundance of several times the solar value, assuming that the region where the emission originates is part of a steady outflow. Since such high Al abundances are very unlikely in the observed objects, a more probable explanation involves more complex density and velocity structures, i.e., layers which are located below the wind acceleration zone and affected by pulsation-induced shocks.

\citet[][]{khouri15} presented a detailed study of the composition and structure of the dusty circumstellar envelope of W~Hya (a well-observed SRa-type AGB star with a relatively low mass loss rate and wind velocity). Their semi-empirical radiative transfer model is constrained by diverse types of observations available in the literature, i.e., scattered light fractions close to the star \citep[][]{norris12}, ISO spectra \citep[][]{sloan03b,just04} and the inner radius of the silicate emission \citep[][]{zhao11}, as well as a gas-phase model \citep[][]{khouri14a,khouri14b}. They conclude that the mid-IR Al$_2$O$_3$ emission is mostly produced in a gravitationally bound dust shell located at about 2 stellar radii, i.e. a region of high density where the precursors of the wind-driving grains form, but where the acceleration of the outflow has not started yet. 
In a paper about scattered light observed around R Car and RR Sco \citet[][]{ireland05} reach similar conclusions, i.e. that the light-scattering dust in the close vicinity of the stars is not part of an outflow. Since the observations are made at shorter wavelengths (around 900 nm), however, the authors can only speculate about the chemical composition of the grains. Taking radiative heating into consideration, they suggest Al$_2$O$_3$ or Fe-poor silicates as possible candidates. 

The scenario discussed by \citet[][]{ireland05} and \citet[][]{khouri15} fits well with the new DARWIN models. As a typical example, Fig.~\ref{f_cmg} shows the variable radial structure of the {\em core-mantle grain} model An315u3cmg with stellar parameters and wind properties that are roughly comparable to W Hya (note also that the resulting variations of visual and near-IR colors are in good agreement with observations of RR Sco and R Car, as shown in Fig.~\ref{f_phot_cmg}). Before silicate condensation sets in at about 2 stellar radii, more than half of the available Al has condensed into Al$_2$O$_3$ grains with radii of about 0.1 micron. The relative strength of radiative to gravitational force in this region is below 10 \%, which makes the term {\em gravitationally bound dust shell} very appropriate. Furthermore, the pulsation-induced shock waves which regularly propagate through these layers lead to a highly variable density structure, with steep gradients due to compression in the shock fronts which may enhance the impression of a shell with a well-defined outer edge. When the silicate mantles of the dust grains start to grow around 2 stellar radii, the radiative pressure increases dramatically, and the acceleration of the outflow starts (note in this context that the ratio of radiative to gravitational forces varies strongly with luminosity, i.e. pulsation phase, spanning a range of about 0.5 to 1.5 in the wind acceleration zone). In this region a transition in the density structure is starting, from the steep atmospheric decline (locally modified by propagating shocks) to a slower decrease of density with distance in the outflow which eventually approaches the $1/r^2$ decline of a wind with constant velocity.

\subsection{The interplay of {Al$_2$O$_3$} and silicate dust}

When comparing our wind-forming {\em core-mantle grain} models (see Tab.~\ref{t_mod_cm}) and the wind models where Al$_2$O$_3$ is treated as a separate passive dust species (Tab.~\ref{t_mod}) with the wind-less {\em {Al$_2$O$_3$} only} models  (Tab.~\ref{t_mod_co}), we found interesting trends of the Al condensation fraction with other model properties. Models with moderate winds, in terms of mass loss rate and outflow velocity, show increased Al$_2$O$_3$ formation compared to the corresponding wind-less pulsating atmospheres (obtained by switching off  silicate formation), while the effect is reversed for stronger outflows. This behavior is probably due to the opposing effects of higher density (i.e., higher growth rates) and higher velocity (less time available for grain growth). It may also explain the observed trend that mid-IR features of Al$_2$O$_3$ appear to be more prominent in low mass-loss objects, while silicate features dominate for stars with higher mass loss rates \citep[see, e.g.,][]{LoMaPom00,sloan03a,karo13}.

Unfortunately, this latter point (i.e. if the different types of mid-IR spectra correspond to different dust compositions in the wind, or rather to different grain temperatures and other conditions in the emitting layers) cannot be tested at present. As discussed by \citet[][]{bladh15}, the intensity of the mid-IR silicate features resulting from the models depends strongly on the temperature of the grains, and therefore on the Fe-content of the silicates further out in the wind. Since a gradual enrichment of the silicate grains with Fe is not included in the current models (which focus on the innermost parts of the circumstellar envelope), no fully consistent mid-IR spectra can be produced at this point. We therefore plan to return to this question in a future paper. 

In this context, however, it is worth to mention some preliminary tests which indicate that thin mantles of Fe-rich silicates on top of Fe-free wind-driving grains (corresponding to a few percent of the total grain radius) may be thermally stable at distances of about 4-5 stellar radii and lead to clearly visible mid-IR silicate features \citep[see][Figs. 9 and 10]{bladh15}. These distances fit remarkably well with silicate condensation radii derived from spectro-interferometric measurements \citep[see, e.g.,][and references therein]{karo13}. A possible interpretation is that the observed spatial gap between the Al$_2$O$_3$ condensation zone (at about 2 stellar radii) and the region where pronounced mid-IR silicate emission originates (at 4 or more stellar radii) corresponds to a region dominated by Fe-free silicate dust. Spectro-interferometric observations of RT Vir by \citet[][]{sacu13} show indeed signs of a weak silicate feature at about 2 stellar radii, implying the presence of Fe-free silicate grains.

\section{Summary and conclusions}\label{s_conclusions}

We have produced new dynamic atmosphere and wind models, in order to study the formation of Al$_2$O$_3$ and silicate dust in the close vicinity of M-type AGB stars, and the resulting effects on mass loss. The equations describing time-dependent grain growth are solved in the framework of a radiation-hydrodynamical model of the atmosphere \& wind, including frequency-dependent radiative transfer for the gas and dust, as well as the effects of pulsation-induced shock waves and luminosity variations. 

In contrast to our earlier papers \citep[][]{hoefner08,bladh15} the models presented here are based on revised growth rates for Fe-free silicate grains, taking into account that the addition of 2 Mg atoms will be the rate-determining step for building Mg$_2$SiO$_4$ monomers in a solar element mixture. The somewhat lower growth rates lead to less efficient wind acceleration and, consequently, to lower wind velocities and mass loss rates than in our earlier models. The basic wind-driving mechanism, radiation pressure due to photon scattering on Fe-free silicate grains, however, remains unchanged. The combinations of wind velocities and mass loss rates resulting from the new models are in good agreement with observations, and the same is true for the visual and near-IR colors, and their variations due to pulsation. 

Regarding the formation of Al$_2$O$_3$ we draw the following conclusions: To make the condensation of this dust species possible at the close distances and in the high concentrations implied by observations \citep[e.g.][]{witt07,zhao11,zhao12,karo13}, the grains have to be quite transparent at visual and near-IR wavelengths. To avoid destruction by radiative heating the value of $k$ (i.e. the imaginary part of the refractive index) should be about $10^{-3}$, or less. This value is similar to lab measurements of a comparable material, i.e. spinel (MgAl$_2$O$_4$)  with Cr impurities at levels compatible with cosmic abundances \citep[a few percent; see][]{zeidler11}. 

Due to the low abundance of Al in a solar mixture, the radiation pressure of Al$_2$O$_3$ (whether from true absorption or scattering) is too low to affect the atmospheric structure significantly, or to drive an outflow for typical AGB stars. This fits well with the scenario of Al$_2$O$_3$ forming a dense, gravitationally bound dust shell at less than 2 stellar radii, as discussed by \citet[][]{ireland05} and \citet[][]{khouri15}. 

Despite its low radiative pressure, Al$_2$O$_3$ may have indirect effects on mass loss: If silicates form as mantles on Al$_2$O$_3$ cores, this may speed up grain growth to sizes relevant for wind driving considerably. Our experimental dynamical models based on such core-mantle grains tend to show higher wind velocities and mass loss rates than models where the outflow is driven by pure silicate grains. Furthermore, the core-mantle grain models lead to variations of visual and near-IR colors during a pulsation cycle which are in even better agreement with observations. 

Finally, we find an intricate feedback mechanism between mass loss and Al$_2$O$_3$ formation: While a moderate wind seems to be beneficial for the condensation of Al$_2$O$_3$ (leading to a higher fraction of Al forming dust than in a wind-less atmosphere), our models show the opposite trend for stronger outflows. It remains to be seen, however, if this mechanism explains 
why mid-IR features of Al$_2$O$_3$ are more prominent in low mass-loss stars, whereas silicate features tend to be more pronounced at higher mass loss rates 
\citep[e.g.][]{LoMaPom00,sloan03a,karo13}.

In summary, the DARWIN models lead to the following picture, which is in good agreement with observations: Al$_2$O$_3$ condenses at distances closer than about 2 stellar radii, forming a gravitationally bound shell, as part of the extended atmospheric layers. Around 2-3 stellar radii, Fe-free silicates start to condense, possibly as mantles on Al$_2$O$_3$ cores. When the grains reach sizes large enough to drive a wind by photon scattering (about 0.1--1 micron) an outflow is triggered. The formation of composite grains with a Al$_2$O$_3$ core and a silicate mantle may give grain growth a head start, increasing both mass loss rates and wind velocities. 
Further out in the wind, the silicates may be gradually enriched with Fe, affecting grain temperature and the intensity of the mid-IR silicate features. 

To take the models from an exploratory to a predictive level regarding Al$_2$O$_3$ formation and mass loss, several issues need to be addressed. In particular, this concerns the intrinsically uncertain visual and near-IR optical properties of circumstellar Al$_2$O$_3$ (microscopic structure and impurities), and the question if Al$_2$O$_3$ and silicates form composite (core-mantle) grains, or separate dust particles. These questions will have to be solved by a combination of theoretical, observational and laboratory studies.

\begin{acknowledgements}
This work has been supported by the Swedish Research Council ({\em Vetenskapsrådet}) and by the ERC Consolidator Grant funding scheme ({\em project STARKEY}, G.A. n.~615604). The computations of spectra and photometry were performed on resources provided by the Swedish National Infrastructure for Computing (SNIC) at UPPMAX. 
\end{acknowledgements}

\begin{appendix} 
\section{RHD equations}\label{app_rhd}

The radiation-hydrodynamical description of the atmosphere \& wind structures is based on
the three conservation laws for mass, momentum and energy, i.e., 
\begin{eqnarray}
  \ppt \rho + \div{(\rho \, u)} & = & 0 \label{eqcont} \\
  \ppt (\rho u) + \div{(\rho u \,u)} & = &  \label{eqmot}\\
     - \grad{\Pg} & - & \frac{G m_r}{r^2} \rho 
     + \frac{4 \pi}{c} \rho \, \left( \kH + \kappa^{d}_{\rm H} \right) H  \nonumber \\
  \ppt (\rho e) + \div{(\rho e \,u )} & = & 
    - \Pg \, \div{u} + 4 \pi \rho \, (\kJ J - \kS S_{\rm g}) \label{eqene}
\end{eqnarray}
(see Table~\ref{t:symbols} for an explanation of the symbols).
The terms on the r.h.s of the equation of motion, i.e. Eq.~(\ref{eqmot}), represent forces due to gas pressure gradients, gravity and radiative pressure on gas and dust, respectively. 
The integrated mass inside a radius $r$ (used to compute the gravitational acceleration) is defined as
\begin{equation}
	m_r = \int_{0}^{r} 4 \pi r'^2 \rho \, dr'  \, .  \label{eqmass}
\end{equation}
The internal energy equation for the gas, i.e. Eq.~(\ref{eqene}), takes changes due to compression or expansion, as well as radiative heating and cooling into account. 
The source terms describing the net energy and momentum exchange
between the gas and the radiation field are frequency integrals, i.e.
\begin{eqnarray}
   \kJ J - \kS S_{\rm g} &=& 
     \int_{0}^{\infty} \kappa^{g}_{\nu} \, ( J_{\nu} - S_{\nu}^{\rm \, gas} ) \, d\nu  \\
   \kH H &=& \int_{0}^{\infty} \kappa^{g}_{\nu} H_{\nu} \, d\nu \\ 
   \kappa^{d}_{\rm H} H &=& \int_{0}^{\infty} \kappa^{d}_{\nu} H_{\nu} \, d\nu 
\end{eqnarray}
where the frequency-averaged 
opacities are defined by
\begin{equation}
  \kappa^{g}_{\rm X} = \frac{\int_{\nu}^{} \kappa^{g}_{\nu} X_{\nu}d\nu}
                        {X}     \qquad {\rm and} \qquad
  \kappa^{d}_{\rm X}   = \frac{\int_{\nu}^{} \kappa^{d}_{\nu} X_{\nu}d\nu}
                        {X}  \,.
\end{equation}
Here $X$ corresponds to one of the quantities $J$, $H$ or $S$,
and variables without the subscript $\nu$ denote the 
frequency-integrated values
$  X = \int_{\nu}^{} X_{\nu}d\nu.$

\begin{table}[t]
\caption{List of symbols used in the RHD equations.}
\label{t:symbols}
\begin{tabular}{cl}
\hline
  &  \\
  $r$              & radius coordinate\\
  $t$              & time \\
  & \\
  $m_{\rm r}$      & mass within radius $r$\\
  $\rho$           & gas density \\
  $e$              & specific internal gas energy \\
  $u$              & matter velocity \\ 
  $J$              & zeroth moment of the radiation intensity \\
  $H$              & first moment of the radiation intensity \\
  $K$              & second moment of the radiation intensity \\
  &  \\
  $P_{\rm g}$      & gas pressure \\
  $\Tg$            & gas temperature \\
  $\Tr$            & radiation temperature \\
  $\Td$            & dust grain temperature \\
  $\kappa^{g}_{\nu}$   & mass absorption coefficient of the gas \\
  $\kappa^{d}_{\nu}$     & absorption coefficient of the dust\\
  $S_{\rm g}$      & source function of the gas\\
  &  \\
  $c$              & speed of light \\
  $G$              & constant of gravitation \\
  $\sigma$         & Stefan-Boltzmann constant \\
  & \\
\hline
\end{tabular}
\end{table}

The frequency-integrated moments of the radiation intensity,
$J$ and $H$, are determined by the zeroth and first order 
moment equations of the radiative transfer equation, i.e. 
\begin{eqnarray}
  \frac{1}{c} \ppt J + \frac{1}{c} \div{(J u)} & = &
      - \div{H} 
      - \frac{1}{c} K \, \div{u} + \frac{u}{c} \frac{3K-J}{r} \nonumber \\
  & & - \rho \, (\kJ J - \kS S_{\rm g}) \label{eqsene} \\
  \frac{1}{c} \ppt H + \frac{1}{c} \div{(H u)} & = &
      - \grad{K} - \frac{3K-J}{r} 
      - \frac{1}{c} H \, \grad{u} \nonumber \\
  & & - \rho \left( \kH + \kappa^{d}_{\rm H} \right) H  \, , \label{eqsflx}
\end{eqnarray}
which are solved simultaneously with the conservations laws for the gas, Eqs.~(\ref{eqcont}) -- (\ref{eqene}).
If the flow velocities are small compared to the speed of light 
(which is the case in AGB star atmospheres and winds) the terms of order $u/c$
in these equations are negligible and the equations take the more 
familiar form
\begin{eqnarray}
  \div{H} + \rho \, (\kJ J - \kS S_{\rm g}) & = & 0  \nonumber\\
  \grad{K} + \frac{3K-J}{r}  + \rho \left( \kH + \kappa^{d}_{\rm H} \right) H & = & 0 \, .  \nonumber
\end{eqnarray}

Considering the energy budget of the dust component, we note that the collisional transfer of energy between gas and dust seems to be negligible compared to the radiative heating and cooling of the dust grains, even for quite transparent grains like Fe-free silicates \citep[see, e.g.,][]{GGS90,bladh15}. Therefore the grain temperature is computed from the condition of radiative equilibrium, i.e.
\begin{equation}
  \kappa^{d}_{\rm J} \, J - \kappa^{d}_{\rm S} \,S \,(\Td) \, = \, 0  \, \, \longrightarrow \, \,
     \Td = \left ( \frac{\kappa^{d}_{\rm J}}{\kappa^{d}_{\rm S}} \right ) ^{1/4} \Tr
\end{equation}
where $S(\Td) = B(\Td) = \Td^4 \sigma / \pi $ 
and the radiation temperature is defined as 
\begin{equation}\label{eq_trad}
\Tr = \left ( J \pi / \sigma \right ) ^{1/4} \,.
\end{equation}

To calculate the frequency-averaged gas and dust opacities which appear in the radiation-hydrodynamical equations and the radiative equilibrium condition for the dust, 
the frequency-depended moments of the radiative intensity, $J_{\nu}$ and $H_{\nu}$,  
have to be known. These moments as well as the Eddington factor 
$  f_{\rm edd} = K / J $
which is needed to close the system of moment equations 
Eq.~(\ref{eqsene}) -- (\ref{eqsflx})
are obtained by solving the frequency-dependent equation 
of radiative transfer for the current density-temperature structure
at each time-step, using the method of characteristics.
Assuming LTE, the source function is given by
\begin{equation}
  S_{\nu} = \frac{\kappa^{g}_{\nu} B_{\nu}(\Tg) + \kappa^{d}_{\nu} B_{\nu}(\Td)}
                {\kappa^{g}_{\nu} + \kappa^{d}_{\nu}}
\end{equation}
where $B_{\nu} (T)$ denotes the Planck function for a temperature $T$.

\section{Variable boundary conditions}\label{app_inbc}

Stellar pulsations play a crucial role for the mass loss of AGB stars by triggering shock waves in the stellar atmospheres which lift gas to distances where dust can condense. 
Since the DARWIN models do not cover the driving zone of the pulsation, the effects of pulsation on the atmosphere \& wind are simulated by prescribing temporal variations of the gas velocity and luminosity at the inner boundary of the models, just below the stellar photosphere.
The most widely used form of this so-called piston boundary \citep[cf.][]{B88} is a simple periodic variation of the gas velocity, 
\begin{equation}
   u_{\rm in} (t) = \Delta u_p \, \cos \left( \frac{2 \pi}{P} \, t \right)
\end{equation}
where $P$ is the pulsation period and $\Delta u_p$ the velocity amplitude at the inner boundary. 
This corresponds to a radial variation 
\begin{equation}
   R_{\rm in} (t) = R_0 + \frac{\Delta u_p P}{2 \pi} \, \sin \left( \frac{2 \pi}{P} \, t \right) 
\end{equation}
simulating the radial expansion and contraction of the pulsating stellar interior. In the current DARWIN models, the accompanying variation of the luminosity $L_{\rm in} (t)$ is parameterized as
\begin{equation}
   \frac{L_{\rm in} (t) - L_0}{L_0} = f_L \, \left(  \frac{R_{\rm in}^2 ( t ) - R_0^2}{R_0^2}  \right)
\end{equation}
where the parameter $f_L$ can be used to adjust the relative amplitudes of velocity and luminosity variations. Note that setting $f_L = 1$ leads to $L_{\rm in} (t) \propto R_{\rm in}^2$, i.e. a constant radiative flux at the innermost layer (but not a constant luminosity) which was used in early models to minimize the number of model parameters. 

\citet[][]{frey08} tested non-sinusoidal variations of $R_{\rm in}$ (derived from 3D star-in-a-box models) and concluded that the critical parameter for triggering a dust-driven wind is the amplitude of the variation, not the shape, since this information gets lost in the transformation of pulsation-induced sound waves into shocks.
In contrast, the shape of the luminosity variation, or a phase shift between $L_{\rm in}$ and $R_{\rm in}$ (which affects the timing of shock propagation through the atmosphere relative to the luminosity variation), may affect the wind properties as recently demonstrated by \citet[][]{liljegren16}. The DARWIN code therefore offers the possibility to specify both $R_{\rm in}$ and $L_{\rm in}$ in terms of Fourier series, to allow for other types of boundary conditions than those specified above.

\section{Numerical methods}\label{app_num}

The conservation laws for the coupled system of gas, dust and radiation form a non-linear system of PDEs which is solved implicitly using a Newton-Raphson scheme. The main features of the numerical technique used in our model calculations are the following: a conservative (volume-integrated) formulation of the discretized radiation-hydrodynamics and dust equations, a monotonic second-order
advection scheme \citep[]{vl77}, as well as an adaptive grid. A so-called grid equation is solved simultaneously with the physical conservation equations and distributes the grid points according to accuracy considerations \citep[]{dd87}. The calculations presented here use 100 radial grid points and the desired resolution is defined by $\rho$~and~$e$, in order to resolve the steep density and temperature gradients in the inner atmosphere and in shock waves.

\end{appendix}

\bibliographystyle{aa} 
\bibliography{hoefner}

\end{document}